\documentclass[fleqn,usenatbib]{mnras}
\usepackage{newtxtext,newtxmath}
\usepackage[T1]{fontenc}
\usepackage{ae,aecompl}
\usepackage{graphicx}	
\usepackage{amsmath}	
\usepackage{amssymb}	

\newcommand{\Msol}{M$_\odot$}
\newcommand{\Teff}{$T_{\mathrm{eff}}$}
\newcommand{\Feh}{\mbox{$\mbox{[Fe/H]}$}}
\newcommand{\Mh}{\mbox{$\mbox{[M/H]}$}}

\title[Young LMC clusters]{Young LMC clusters: the role of red supergiants and multiple stellar populations in their integrated light and CMDs}

\author[Asa'd et al.]{Randa S. Asa'd,$^{1}$\thanks{E-mail:raasad@aus.edu}
Alexandre Vazdekis,$^{2,3}$
Miguel Cervi\~no,$^{2,3,4}$
\newauthor
Noelia E. D. No\"el,$^{5}$
Michael A. Beasley,$^{2,3}$
Mahmoud Kassab,$^{1}$
\\
$^{1}$American University of Sharjah, Physics Department, P.O.Box 26666, Sharjah, UAE\\
$^{2}$Instituto de Astrof\'isica de Canarias (IAC), E-38200 La Laguna, Tenerife\\
$^{3}$Departamento de Astrof\'isica, Universidad de La Laguna, E-38205 Tenerife, Spain\\
$^{4}$Instituto de Astrof\'isica de Andaluc\'ia (IAA-CSIC), Glorieta de la Astronom\'ia s/n, 18008 Granada, Spain\\
$^{5}$Department of Physics, University of Surrey, Guildford GU2 7XH, UK
}

\date{Accepted XXX. Received YYY; in original form ZZZ}
\pubyear{2017}

\begin{document}

\label{firstpage}
\pagerange{\pageref{firstpage}--\pageref{lastpage}}
\maketitle

\begin{abstract}
 The optical integrated spectra of three LMC young stellar clusters (NGC~1984, NGC~1994 and NGC~2011) exhibit concave continua and prominent molecular bands which deviate significantly from the predictions of single stellar population (SSP) models.
   In order to understand the appearance of these spectra, we create a set of young stellar population (MILES) models, which we make available to the community.
 We use archival {\it International Ultraviolet Explorer} integrated UV spectra to independently constrain the cluster masses and extinction, and rule out strong stochastic effects in the optical spectra. In addition, we also analyze deep colour-magnitude diagrams of the clusters to provide independent age determinations based on isochrone fitting. 
   We explore hypotheses including age-spreads in the clusters, a top-heavy initial mass function, different SSP models and the role of red supergiant stars (RSG). We find that the strong molecular features in the optical spectra can only be reproduced by modeling an increased fraction of about $\sim20$ per cent by luminosity of RSG above what is predicted by canonical stellar evolution models. Given the uncertainties in stellar evolution at Myr ages, we cannot presently rule-out the presence of Myr age-spreads in these clusters.
   Our work combines different wavelengths as well as different approaches (resolved data as well as integrated spectra for the same sample) in order to reveal the complete picture. We show that each approach provides important information but in combination can we better understand the cluster stellar populations.

\end{abstract}  
 

 \section{Introduction}
 
 One of the most significant results of the past decade in the field of star clusters is the discovery of extended main-sequence turn offs (eMSTO) of intermediate-age star clusters (100 Myr $<$ age $<$ 10 Gyr) in the Large Magellanic Cloud (LMC). Perhaps even more intriguing are the recent serendipitous findings of eMSTOs in very young stellar clusters (age $<$ 100 Myr) as well. The origin of eMSTOs in young LMC clusters is presently debated in the community \citep{Milone13, Milone15, Milone16, Milone17a, Correnti15, Correnti17a, Bastian13a, Bastian16, Niederhofer15a, Niederhofer15b}. However, if eMSTOs are caused by age spreads within individual clusters, on order of tens of Myrs instead of hundreds of Myrs as in the case of intermediate-age LMC clusters, this would strongly affect our understanding of the formation of stellar clusters typically thought to be simple stellar populations (SSPs). The possible existence of age spreads have been investigated in young massive clusters in other galaxies as well \citep {Larsen11, Bastian13b}.  Most of this analysis of multiple stellar populations (MSPs) has been performed using colour-magnitude diagrams (CMDs) with only one study using integrated spectra \citep {Cabrera-Ziri14}  to explore possible MSPs in LMC clusters.

For very young LMC clusters, those with ages below $\sim$30\,Myr, the most massive stars dominate almost completely the integrated light. In fact, the strong contrast in the spectral shape of the blue
and red supergiants (BSGs, RSGs) determines the shape of the spectrum of the
stellar population, which will be determined by the relative weight between
these stars as a function of wavelength. This is particularly relevant for ages
around $\sim$10\,Myr where the net contribution of the RSGs to the total light
peaks, affecting dramatically not only the spectrum shape but also some molecular
band features (e.g. \citealt{Mayya97}). The extent of this effect will depend
heavily on the Blue to Red Supergiants ratio, which is predicted to decrease
with increasing metallicity (e.g., \citealt{Langer95}), although at odds with
the observational constrains (e.g., \citealt{Eggenberger02}). This ratio, which
depends on mass loss, convection and mixing processes, is therefore key for
testing both the stellar evolutionary models and also the stellar population
models in this age regime.

The LMC constitutes an excellent laboratory to study 
MSP in young star clusters. Its proximity, at only $\sim$50 kpc from us, makes it straightforward to
analyse resolved individual stars in clusters and integrated properties at the same time (e.g., \citealt{Olsen99};
\citealt{Beasley02}).
However, both kinds of studies require careful consideration of the total number of stars each cluster hosts \cite[see discussions in][]{Chiosi88,Chiosi06}.
In this work we study the integrated optical spectra of three young LMC clusters
using a full spectrum-fitting technique in conjunction with UV spectra and CMDs from archival data. 
We aim at studying their relevant
stellar population properties, including the MSP scenario and also testing the
models in this age regime.

Our motivation comes from the fact that when fitting the integrated spectra of these three
young LMC clusters (age $<$ 50 Myr) with different stellar population
models \citep{Asad16}, we noticed a concave continuum shape of the
observational data that prevents a good fit with these models, as well as a mismatch of the molecular features of the observations when compared to SSPs.
The spectra of the young clusters used in this work are
taken from different observing runs and instruments which rules out possible flux-calibration issues, this effect was noted by others as well (Gelys Trancho, private communication). 
Here we investigate if the mismatch has the same origin as the split noticed in the MSTO of older
clusters or whether these are to be attributed to current limitations such as the
relative contributions of Blue and Red supergiants. 

In section \ref{Sec_data} we describe the cluster sample, the data used in this work, and previous results. In section \ref{Sec_models} we describe the newly computed young MILES models used in this work. The results obtained with standard SSP fitting method is presented in section \ref{Sec_1SSP}. Our analysis for improving the fits are discussed in section \ref{Sec_2SSP}. We present the affects of varying the RSG contribution in section \ref{Sec_RSGs}.
In section \ref{Sec_checking} we perform a cross check using 
UV data and a CMD analysis.
Our summary and discussion is given in \ref{Sec_discussion}.

 \section{Data Sample and previous results}
\label{Sec_data}

\begin{table}
\caption{IUE files used in this work.}
\label{tab:iue}
\begin{tabular}{rcc}
\hline
Name & SWP (Exp. Time [s])& LWP (Exp. Time [s]) \\ 
\hline
NGC~1984 & 23252L (419) & 03564L (269)\\
NGC~1994 & 23254L (539) & 03566L (269)\\
NGC~2011 & 39260L (2699)  & 18402L (1799) \\
\hline
\end{tabular}
\end{table}

We present the analysis of three young clusters NGC~1984, NGC~1994 and NGC~2011 in
the optical range 3800-6230\,\AA, together with complementary analysis in the UV and a CMD analysis. We chose these three young clusters from our previous work \citet{Asad16} which analyzed a wider age range. For the purpose of this work we picked the young clusters with available CMDs.

\subsection{Optical data}
\label{Sec_data_optical}

Optical data was obtained 
in two observing
runs in 2011 with the RC spectrograph on the 4 m Blanco telescope (NGC~1984 and
NGC~2011 with resolution 14\,\AA) and with the Goodman spectrograph on the SOAR
telescope (NGC~1994 with resolution 3.6\,\AA). We obtained integrated spectra by scanning
the cluster S-N, with the slit aligned east-west.  As results of this process we obtained the spectra of the central region of the clusters, covering diameters of  26'' ($r = 0.2'$), 18'' ($r = 0.15'$) and 16'' ($r = 0.13'$) for NGC~1984, NGC~1994, and NGC~2011 respectively.

 \subsection{UV data}
\label{Sec_data_UV}

In addition, we have used archival {\it International Ultraviolet Explorer} (IUE) UV spectra of the three clusters provided by the INES data center\footnote{\tt http://sdc.cab.inta-csic.es/ines/} \cite[and references therein]{GRCW01} . The IUE data summary we use is shown in Table~\ref{tab:iue}. Where several spectra were available, we tested the headers of the FITS files and the self consistency of the data, and chose those with the highest signal to noise, least number of bad pixels and best agreement in the overlapping region between SWP and LWP spectra. All the IUE spectra where obtained with the large 10x20 arcsec aperture, which is similar to the aperture of our optical spectra.

\subsection{CMD data}
\label{Sec_data_CMD}

We have also analyzed the CMDs for all three clusters using data from the Survey of the
Magellanic Stellar History \cite[SMASH,][]{Nidever2017}.  SMASH is a
NOAO\footnote{NOAO: National Optical Astronomy Observatory
https://www.noao.edu/} community DECam survey of the Magellanic Clouds mapping
480 deg$^2$ to  24th magnitude in {\it ugriz} filters  with the goal of identifying
broadly distributed, low surface brightness stellar populations associated with
the stellar halos and tidal debris of the Magellanic Clouds. The pipeline used
to process the data from SMASH performs a point spread function (PSF) photometry
based on the DAOPHOT software suite \citep{Stetson87}.

\subsection{Previous analysis of the clusters}
\label{Sec_data_previous}

The present set of clusters have been analyzed using different wavelength coverages and methodologies. These clusters have been also analyzed in the UV.  Studies based on CMD analysis \cite[e.g.][]{2010Glatt} found ages around 50 Myr with a typical error of 0.3 dex and studies based on integrated colours obtained much younger ages (typically younger than 10 Myr, see a summary of results in \citealt{asad12}) with the exception of  \cite{Popescu2012} who obtained ages similar to those obtained from CMDs analysis. In \cite{Popescu2012}, the authors infer cluster masses a few times $10^4$M$_\odot$ \cite[masses also obtained in][]{asad12}, they apply Montecarlo simulations to take into account the fluctuations in the cluster luminosities due to their relatively low masses.  \cite{VGetal2017} obtain ages, metallicities and mass estimates of NGC~1994 and NGC~2011 by the analysis of UV spectral features to estimate age and metallicities and V-band photometry to estimate masses using a fitting to both, Monte Carlo simulations and analytical modeling; they found ages about $19$ and $11$ Myr, and masses about $1.1 - 1.7 \times 10^4$ and about $3-8 \times 10^3$ M$_\odot$ for NGC~1994 and NGC~2011 respectively; in both cases their provide metallicities about 0.015 as their best fitted results. 

\cite{Asad16} used integrated spectra and found younger ages, in some case between 3 and 4.4 Myr with $E(B-V)$ values larger than 4.4 \citep{asad12}. With the UV data, NGC 1984 and NGC 1994 has been analyzed by \cite{CBG87} which obtain total $E(B-V)$ values of 0.14 for both clusters including   the foreground colour excess 
 due to our Galaxy. They obtain from both clusters that the UV light profile is mainly concentrated in the core of the cluster although showing extended asymmetric wings. 
NGC 2011 has been analyzed by \cite{KKM93} as part of a study of binary LMC clusters and show that it is part of a three component system, with a dynamical mass of the main component  of $\sim5 \times 10^5 \mathrm{M}_\odot$. In Sect.~\ref{Sec_sampling} we discuss more in details Mass estimates by different authors.

These previous results show that the evolutionary status of those clusters is not clear and depends on the wavelength range, methodology, and, possibly, the aperture used for the analysis. Therefore, one of our goals is also to understand the reasons of those differences.

For the present study we assume a distance modulus  of $(m - M) = 18.50$ mag \cite[e.g.][]{Alves2004}, which sets the LMC at $\sim$50 Kpc from us. We adopt a Galactic extinction in the LMC direction of $E(B-V)_\mathrm{G} = 0.06$, obtained from the $A_\mathrm{V}=0.20$ value from \cite{SF11} as given in NED\footnote{\tt https://ned.ipac.caltech.edu/}.

\section{MILES stellar population models with young ages}
\label{Sec_models}

\subsection{SSPs}
\label{Sec_SSP}

\begin{figure}
\resizebox{75mm}{!}{\includegraphics[angle=270]{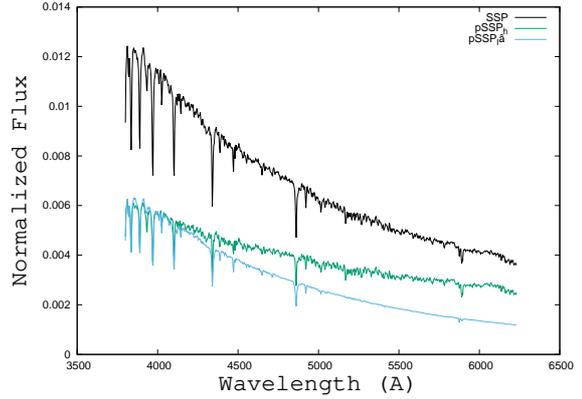}}
\caption{Partial SSP spectra corresponding to an age of 10Myr and LMC
metallicity computed with the stars with masses below (blue)
and above (green) $17$\,M$_{\odot}$. The two pSSPs added provide the SSP (black).
There is a significant change in the slope with the high mass partial SSP
dominating the light in the red spectral range and all the molecular features
there,whereas the blue spectrum has fewer features than the green one.}
\label{fig:all.eps}
\end{figure}

For the purposes of this study, we extended the MILES single-age, single-metallicity SSP models \citet{Vazdekis10,Vazdekis16} to predict SEDs at resolution
2.5\,\AA\ for ages as young as 6.3\,Myr. The models employ the scaled-solar
stellar evolutionary isochrones of \citet{Bertelli94}, extended to very low mass
stars (M$<0.5$\Msol\ with the stellar tracks of \citet{Pols95} as implemented in
\citet{Vazdekis96}. These models also employ the empirical stellar spectral library
MILES \citep{MILESI}, composed of nearly 1000 stars, mostly from the field in 
the solar neighborhood.
Although there are important differences in the input physics adopted for computing these isochrones with respect to the
\citet{Girardi2000} (used in more recent versions of the MILES models), such as the
equation of state, the opacities or the convective overshoot scheme, they can
be safely matched at ages around 60\,Myr, as shown by these authors.
The theoretical parameters of the isochrones are converted to the observational
plane, i.e. stellar fluxes, on the basis of empirical relations between colours
and stellar parameters (temperature, metallicity and gravity), instead of using
theoretical stellar spectra. We mostly use the metallicity-dependent empirical
relations of \citet{Alonso96, Alonso99}; respectively, for dwarfs and giants.
For stars with temperatures above $\sim8000$\,K we use the empirical compilation
of \citet*{Lejeune97, Lejeune98}. We also use the metal-dependent bolometric
corrections of \citet*{Alonso95} and \citet{Alonso99} for dwarfs and giants,
respectively, and adopt ${\rm BC_{\odot}}=-0.12$. Assuming ${\rm V_\odot}=26.75$
\citep{Hayes85} we obtain an absolute magnitude for the Sun of ${\rm
M_{V_\odot}=4.82}$ and ${\rm M_{{bol}_{\odot}}}$ is given by ${\rm M_{V_{\odot}}+BC_{V_{\odot}}}= 4.70$.

It is worth noting that as the empirical stellar spectra follow the Milky-Way
abundance pattern as a function of metallicity, the resulting models are nearly
consistent and scaled-solar around solar metallicity whereas at low metallicity
they lack consistency. Therefore these models, which assume that $\Mh=\Feh$,
should be identified as "base" models, following the description given in
\citet{Vazdekis15}. 
The newly computed models do not include stellar rotation, which, among other
effects, brings more material to the convective core of massive stars  increasing their MS lifetimes by as much as $\sim25\%$ and decreasing 
the surface gravity and the opacity in the radiative envelope, which in turns raises the
luminosity \citep{Maeder00}. The resulting colours of the stellar
populations with ages smaller than $\sim40$\,Myr vary by $0.1$ -- $1$\,mag with
respect to the non-rotating models. According to \citet{Vazquez07} the effect is
larger with increasing wavelength and metallicity.

An SSP is computed by integrating the stellar spectra along the isochrone. For
each star, characterized by its main atmospheric parameters (\Teff, $\log g$,
\Mh), we use the spectra of the stars of the MILES database with the closest
parameters following the local interpolation scheme described in \citet{CATIV},
as updated in \citet{Vazdekis15}. We scale the stellar spectra according to the
absolute flux in the $V$-band following the method described in
\citet{Falc11}. The number of stars in each mass bin comes from the
adopted Inital Mass Function (IMF), for which we assume five shapes as summarized in \citet{Vazdekis16}. In
this particular work we focus on the low mass tapered (regarded as Bimodal)
IMF, which is characterized by the logarithmic slope, $\Gamma_b$, as a free
parameter. The Kroupa Universal IMF \citep{Kroupa01} is very similar to a bimodal IMF with
$\Gamma_b=1.3$. As reference, for the clusters masses and number of stars obtained in this work, we assume for all the IMFs a lower mass limit of 0.1 M$_\odot$, and the upper mass limit of 100 M$_\odot$.

\subsection{Partial SSPs}
\label{Sec_TestpSSP}

We also created a library of partial MILES SSPs (hereafter pSSPs) for both LMC and Solar
metallicity. These pSSPs are used to test the RSG prescription at the LMC metallicity and the scenario where low mass require longer time scales to land into the zero age main sequence (ZAMS).
We compute pSSPs where only the contributions of stars above a given initial mass $M_p$ are considered (we call it pSSP$_{h}$), and other models were only the stars with masses below $M_p$ are integrated
(pSSP$_{l}$). Note that the addition of the spectra of these two pSSPs gives the
SSP spectrum (pSSP$_{l} + $ pSSP$_{h} = $ SSP). Note also that it is possible to compute pSSPs with the contribution of stars within a given, intermediate, mass range.

Figure~\ref{fig:all.eps} shows two pSSPs with LMC metallicity and 10\,Myr, where M$_p$=17\,M$_{\odot}$. These two pSSPs added match the corresponding SSP. Note that there is a
significant change in the slope with the partial SSP corresponding to the stars
with $M >17$\,M$_{\odot}$ dominating the light in the red and all the molecular features
there, whereas the blue spectrum is featureless.

We use these pSSPs to investigate the possible impact on the resulting SEDs of the fact that
lower-mass stars require much longer time-scales to reach the ZAMS, which can be
as long as $\sim 1$\,Gyr for stars of $\sim 0.1$\,\Msol,  in comparison to their
higher mass counterparts \citep{Bodenheimer11}. By combining pSSPs with varying ages we can derive the SED of a single star formation burst but with stellar mass ranges in different ages. 
The motivation is two-fold: Firstly, it allows us to assess the contribution of stars in different mass ranges (and the RSGs in particular) to the total light, and analyze how the spectra vary with changes in the RSGs evolutionary prescriptions 
(see Section~\ref{Sec_pSSPs_RSGs} for more details). Secondly, it allows us to study the sampling effects (discussed in Section~\ref{Sec_sampling}) for the case where the isochrone might not become fully populated. 
This is performed with  {\sc{sed}@} synthesis code  \cite[][and references therein]{MHK91,Cervin06} using the same IMF and isochrones as in the MILES models, but with the low resolution BaSeL2 
theoretical atmosphere  library from \cite{Lejeune97,Lejeune98} with LMC metallicity.

The LMC metallicity is well covered by the MILES library (see \citealt{Cenarro07} and references therein), including RSGs such as the one shown in Fig. 9. In this analysis we neglect any abundance ratio difference between the LMC and Field stars. However these possible variations have an impact on small wavelength scales.

 \section{Results Using Single Stellar Populations}
 \label{Sec_1SSP}

\begin{table*}
\caption{Results using the Kroupa Universal IMF for several scenarios (c.f. Sects.~\ref{Sec_1SSP} and \ref{Sec_2SSP})}
\label{tab:Results1}
\begin{tabular}{lcccccc}
\hline
Name & Metallicity & Age$^1$ & E(B-V)$^1$ & Age$^2$ & Age$^3$ &  $f_7^4$ \\
 &  & $\log(\mathrm{age/yr})$ &  & $\log(\mathrm{age/yr})$ & $\log(\mathrm{age/yr})$ &   \\
\hline
NGC~1984 & LMC &6.9 & 0.03 & 6.9 & 6.9 & 0.00 \\
NGC~1994 & LMC &7.0 & 0.02 & 7.0 &     &      \\ 
NGC~2011 & LMC &7.0 & 0.10 & 6.9 & 6.9 & 0.30 \\
NGC~1984 & SOL &7.0 & 0.01 & 7.0 & 6.9 & 0.97 \\
NGC~1994 & SOL &6.8 & 0.10 & 7.3 &     &      \\ 
NGC~2011 & SOL &7.2 & 0.02 & 7.2 & 6.8 & 0.92 \\
\hline
\multicolumn{7}{l}{$^1$ Age and E(B-V) using 1SSP with reddening correction. }\\
\multicolumn{7}{l}{$^2$ Age using 1SSP neglecting reddening correction.}\\
\multicolumn{7}{l}{$^3$ Age using 2 SSPs where one of them have a fixed age of 10 Myr according equation \ref{Eq_2SSPs} in Sec.~\ref{Sec_IMF_13}.}\\
\multicolumn{7}{l}{$^4$ Fraction of the contribution of the 2nd SSP according equation \ref{Eq_2SSPs} in Sec.~\ref{Sec_IMF_13} }
\end{tabular} 
\end{table*}

\begin{figure}
\begin{tabular}{ccc}
\resizebox{75mm}{!}{\includegraphics[angle=0]{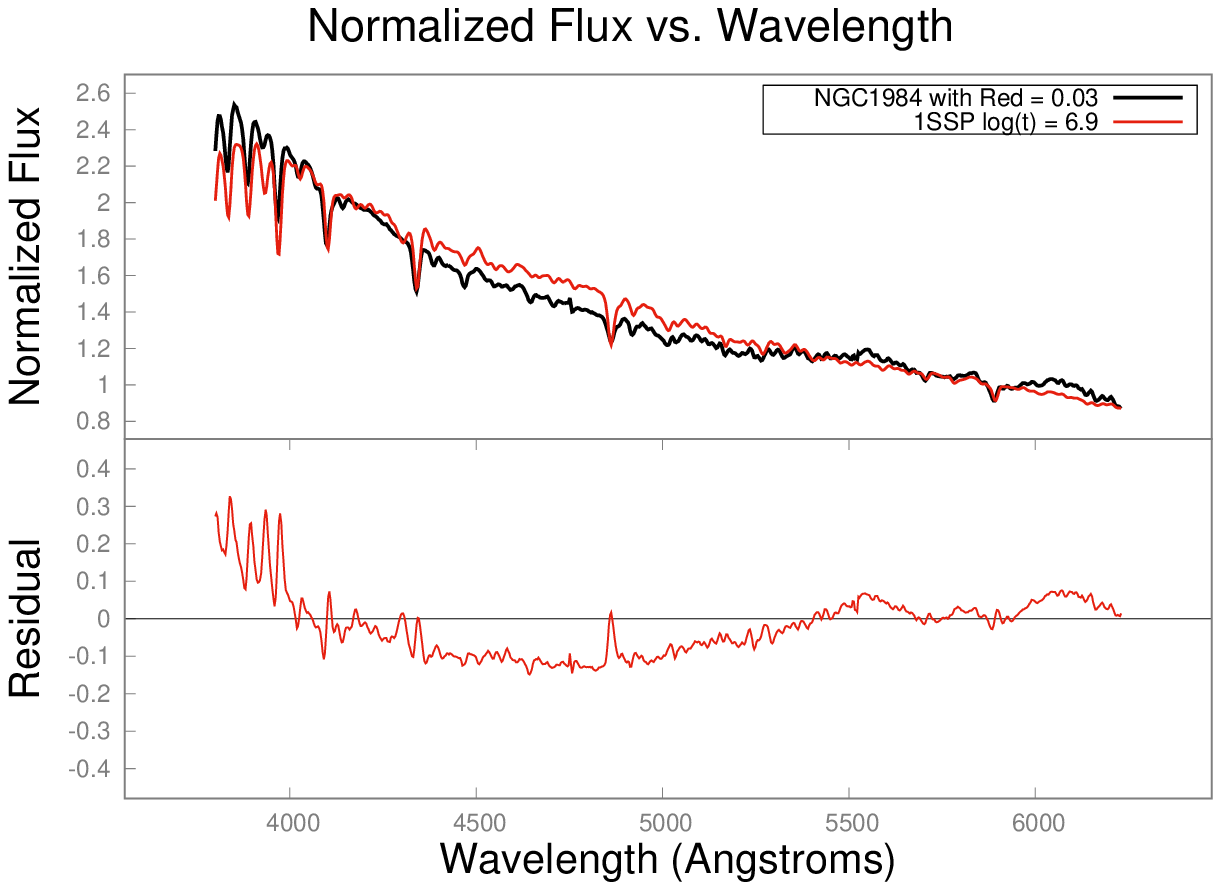}} \\
\resizebox{75mm}{!}{\includegraphics[angle=0]{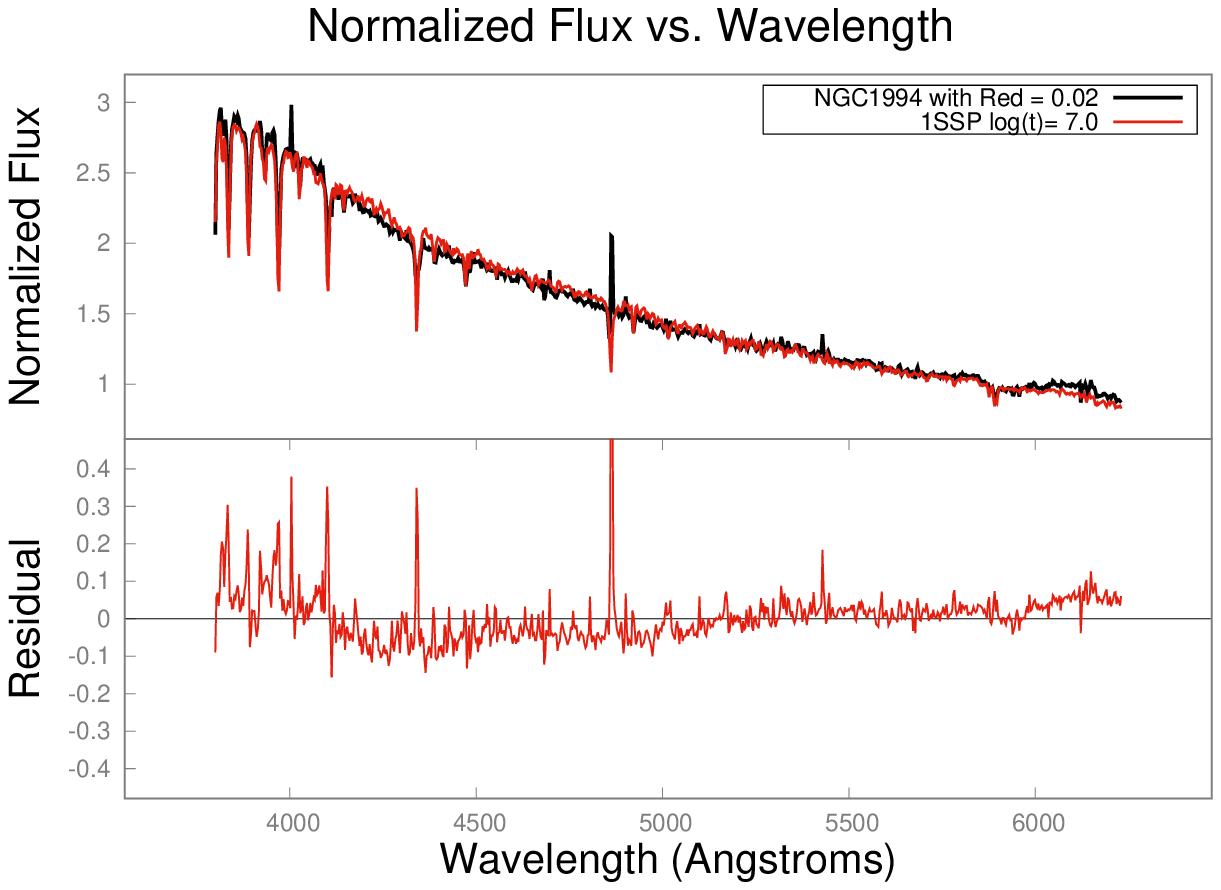}} \\ 
\resizebox{75mm}{!}{\includegraphics[angle=0]{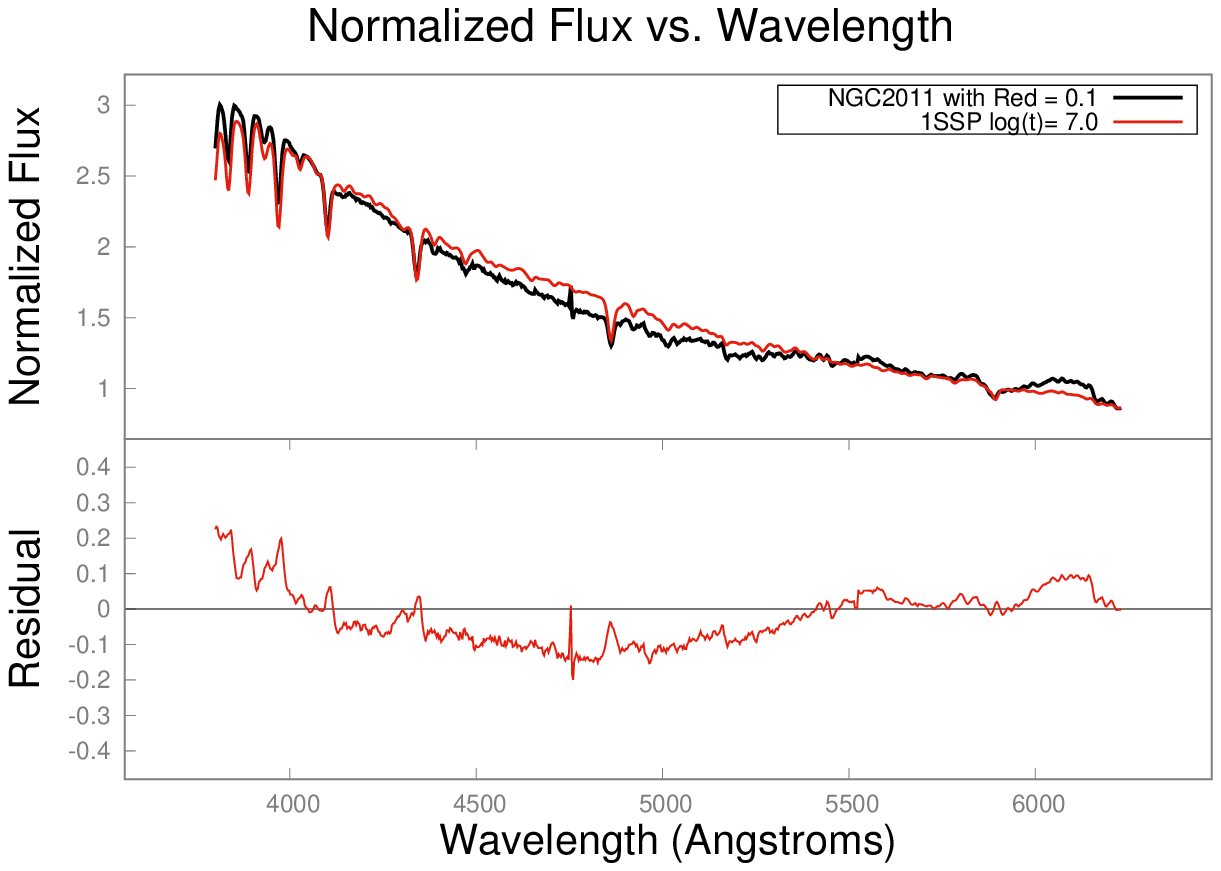}} \\
\end{tabular}
\caption{Best $\chi^2$ fits achieved with 1SSP MILES models using LMC metallicity and Kroupa Universal IMF. The fits correspond to the location of the star symbol in Fig.~\ref{fig:SSP_Prob}.}
\label{fig:SSPred}
\end{figure}

\begin{figure}
\begin{tabular}{ccc}
\resizebox{75mm}{!}{\includegraphics[angle=0]{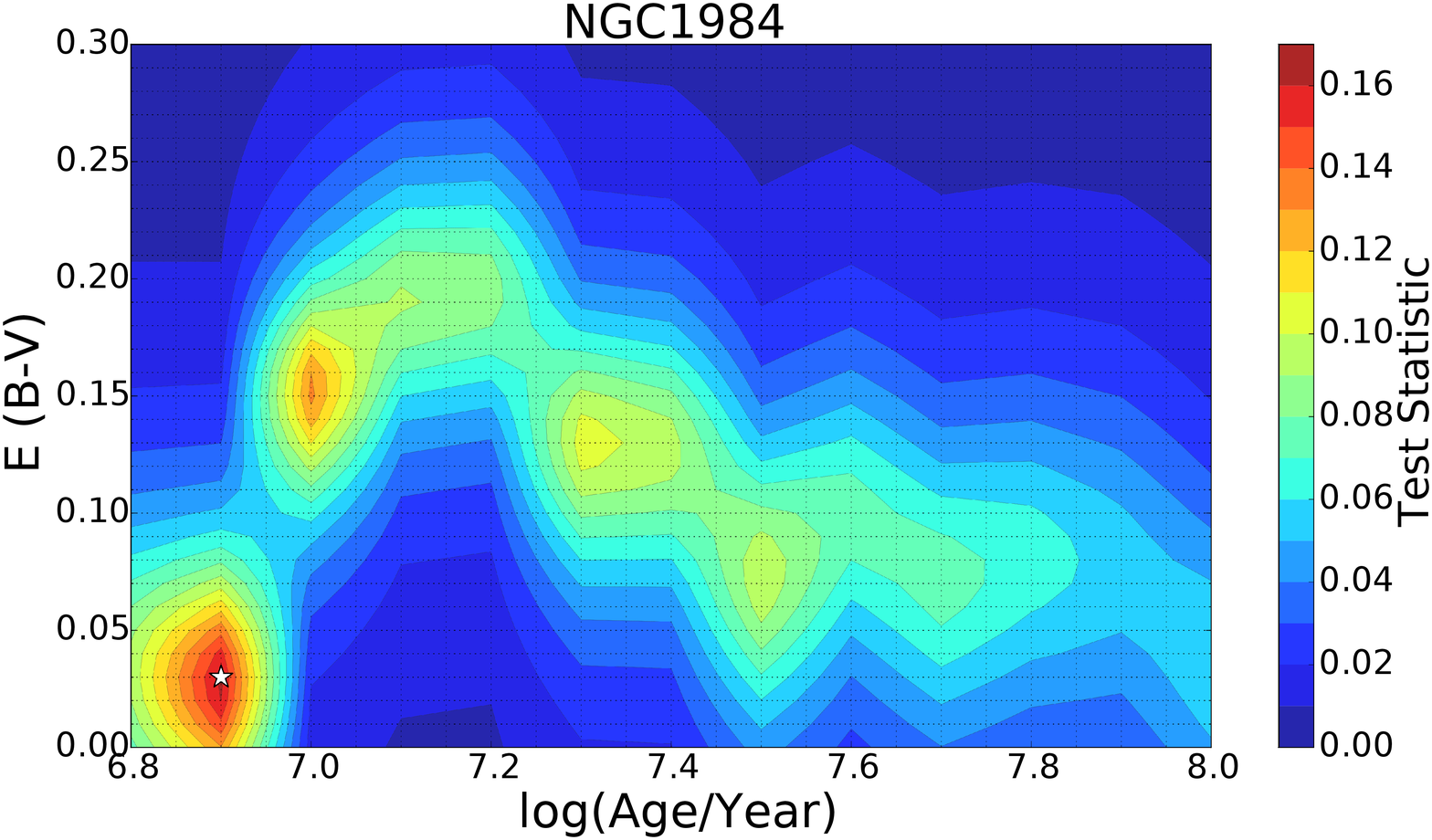}} \\
\resizebox{75mm}{!}{\includegraphics[angle=0]{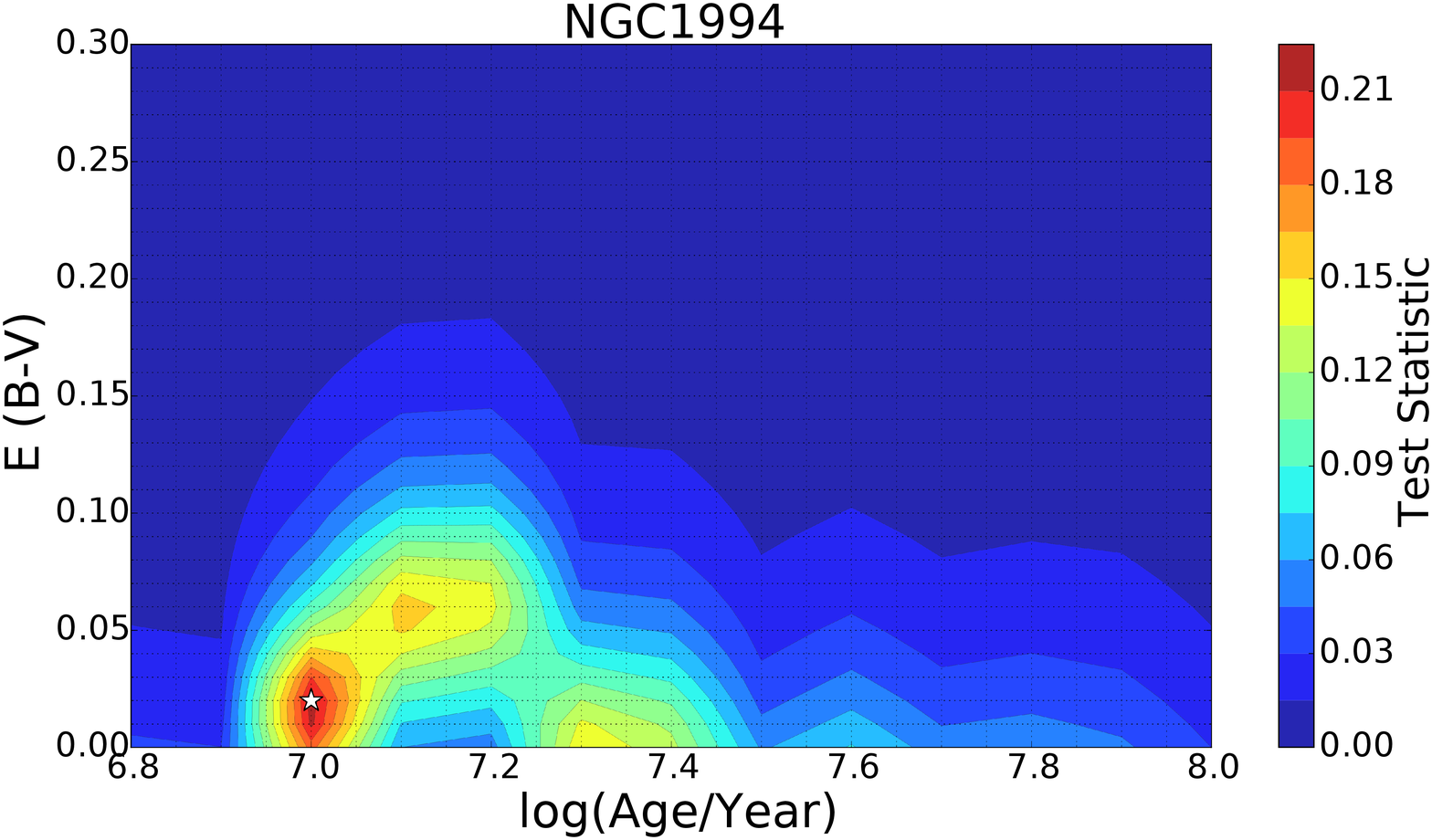}} \\
\resizebox{75mm}{!}{\includegraphics[angle=0]{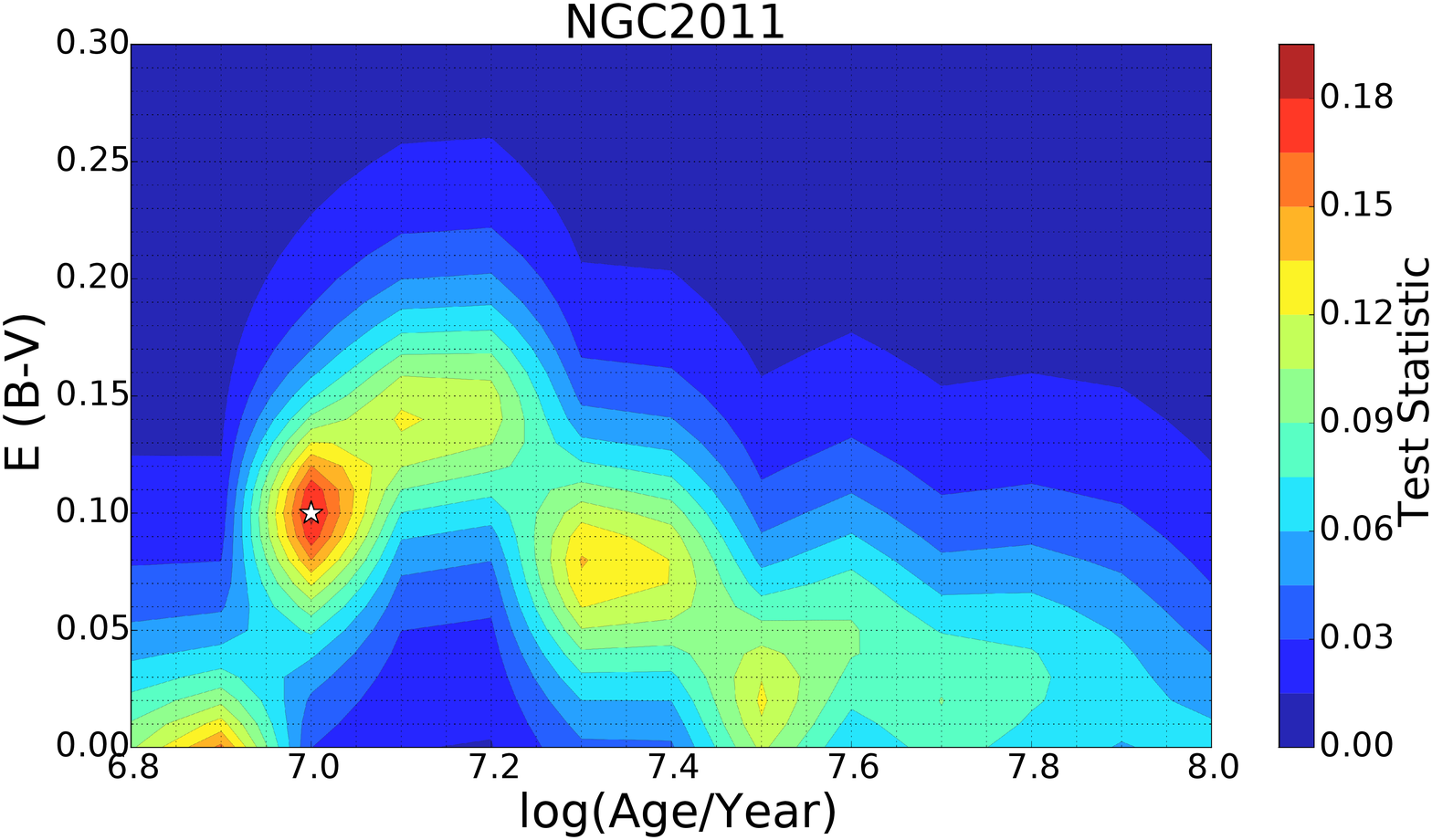}} \\
\end{tabular}
\caption{Probability density plots in the $\log t$ - $E(V-B)$ plane of the 1SSP solutions obtained with MILES models (LMC metallicity and Kroupa Universal IMF). The best $\chi^2$ solution is shown as a star and corresponds to the fit shown in Fig.~\ref{fig:SSPred} The test statistics value is the inverse value of the $\chi^2$ solution}
\label{fig:SSP_Prob}
\end{figure}

We start by obtaining the age of the clusters by comparing SSPs to our observed integrated spectra with the full spectrum-fitting algorithm as described in \citet{asad13,Asad16}.
The \citet{Asad16} analysis used the \cite{GALAXEV} SSP models 
because the MILES models did not include young SSPs at that time. Here we present the results with the newly computed young MILES SSPs. We first perform fits with reddening as a free parameter and the result are listed in the first two columns of Table~\ref{tab:Results1}. Fig.~\ref{fig:SSPred} shows the best  $\chi^2$ matching our observation using the LMC metallicity, Kroupa Universal IMF and MILES models. Figure~\ref{fig:SSP_Prob} shows the probability density plots in the $\log t$ - $E(V-B)$ plane of the obtained solutions. However, no solution using the nominal best $\chi^2$ is compatible with presence of a galactic extinction in the LMC direction of $E(B-V)_\mathrm{G} = 0.06$. For NGC~1984 and NGC~2011 there exists a solution at 10 Myr with $E(B-V) > E(B-V)_\mathrm{G}$. In the case of NGC~1994, a compatible (reddening) solution requires an age of $\log t= 7.1$. We note that, this cluster shows the presence of emission lines in its spectra, which are clearly seen in the residuals of the
spectral fit.

It is evident that the observed and model spectra do not perfectly match. The situation is more extreme for the case of NGC~1984 and NGC~2011, the residuals show a concave-like continuum and a characteristic mismatch above $\sim$5500\,\AA, which can be attributed to strong molecular features (e.g, TiO). 
A similar, although not so extreme curvature is present in NGC~1994. We have tested the residuals obtained by other age and extinction combinations for the three clusters and they also show a similar concave-like pattern. Such behaviour is also present in the results of \citet{Asad16} for clusters with ages around 10 Myr which make use of \cite{GALAXEV} synthesis code \cite[see Figs. 22 and 23 in][for NGC~1994 and NGC~2002, this last one not studied here]{Asad16}.
We tested changing the evolutionary prescriptions by using 
the Geneva-based \citep{Schaerer1993} Granada atmosphere models \citep{Delgado05} computed with {\sc sed}@ synthesis code and, again, similar residuals with a concave continuum is obtained. Hence this problem looks to be generic to evolutionary models and evolutionary prescriptions around an age of 10 Myr, and is not a MILES-model specific problem.

\subsection{Line-strength analysis}
 \label{Sec_indices}

\begin{figure}
\includegraphics[angle=0,scale=0.45]{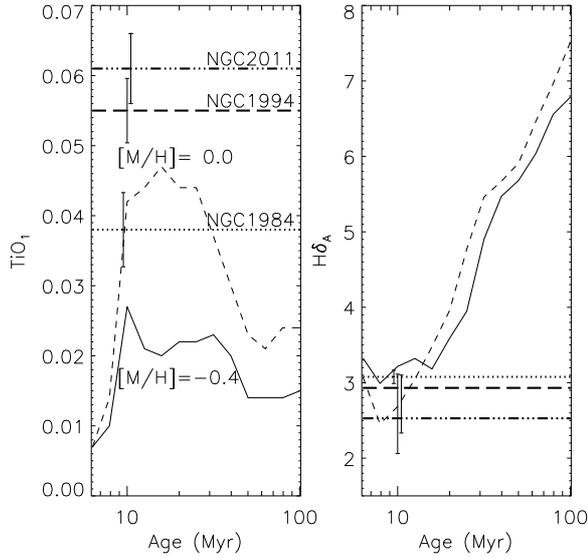}
\caption{Line-strength analysis. The left panel shows the time evolution of the TiO$_1$ molecular feature index for the MILES young SSP models with LMC (solid line) and solar (dashed line) metallicity. The cluster index values are represented by the horizontal thick lines, with different layouts as marked above them. All the spectra were smoothed to match a resolution of 14~\AA (FWHM). Errorbars are shown around 10\,Myr. Note that neither the SSP models nor any combinations of SSPs with LMC metallicity are able to match the observed cluster index values. The right panel shows the higher order Balmer line index H$\delta_{A}$ (wide index definition), which can be considered as an age indicator. All the lines have the same meaning as in the left panel. For NGC\,1994 we used the emission filling-in cleaned spectrum obtained in Section~\ref{Sec_ppxf}. The lower end of the errorbar of this cluster includes the index measurement without applying any emission filling-in correction. No emission filling-in was detected for NGC\,1984, whereas for NGC\,2011, if any, the correction is negligible. From the H$\delta_{A}$ values we can conclude that all the clusters have ages smaller than $\sim$15\,Myr.}
\label{fig:indices}
\end{figure}

\begin{figure}
\resizebox{75mm}{!}{\includegraphics[angle=0]{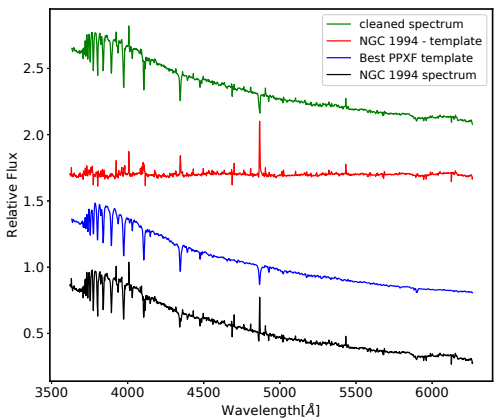}}
\caption{NGC~1994 emission analysis obtained with pPXF and described in Sect. \ref{Sec_ppxf}.}
\label{fig:ppxf}
\end{figure}

\begin{figure}
\begin{tabular}{ccc}
\resizebox{75mm}{!}{\includegraphics[angle=0]{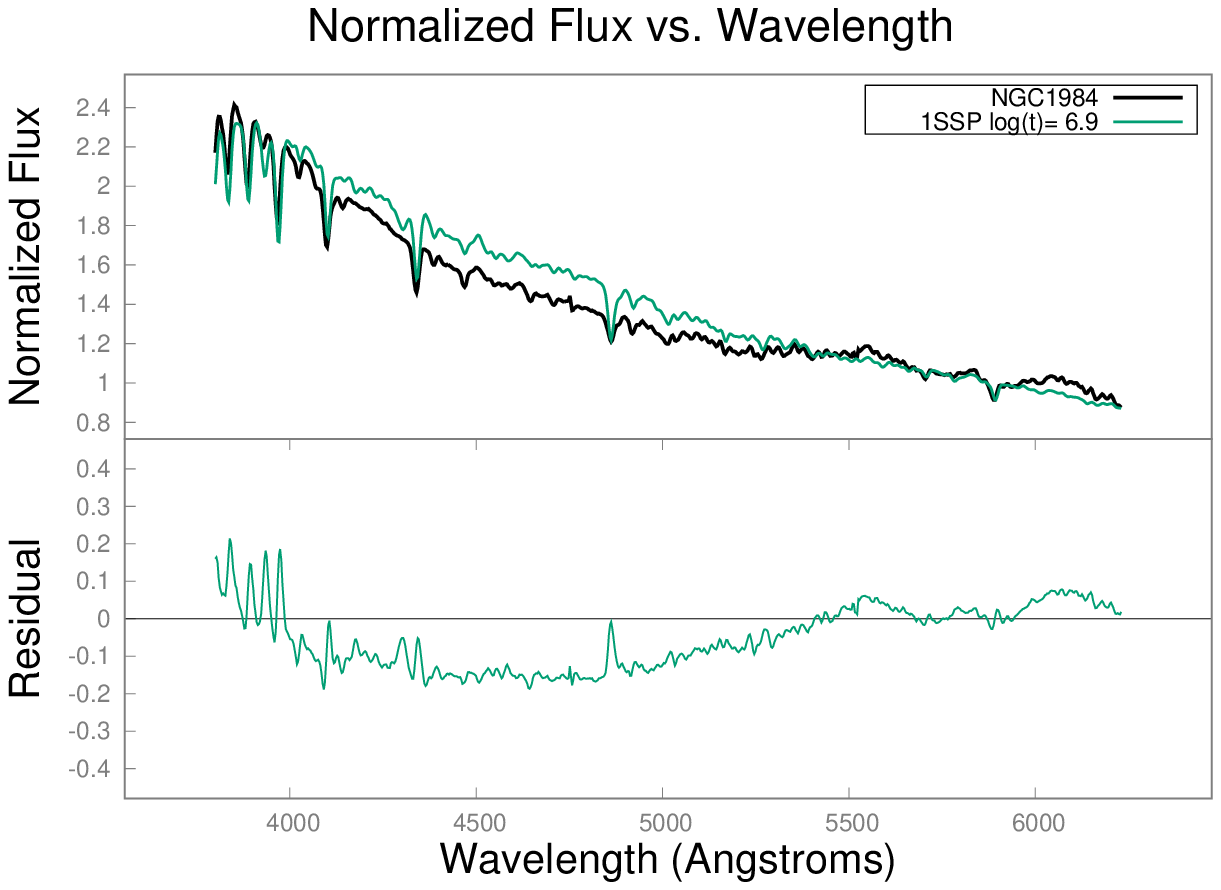}} \\
\resizebox{75mm}{!}{\includegraphics[angle=0]{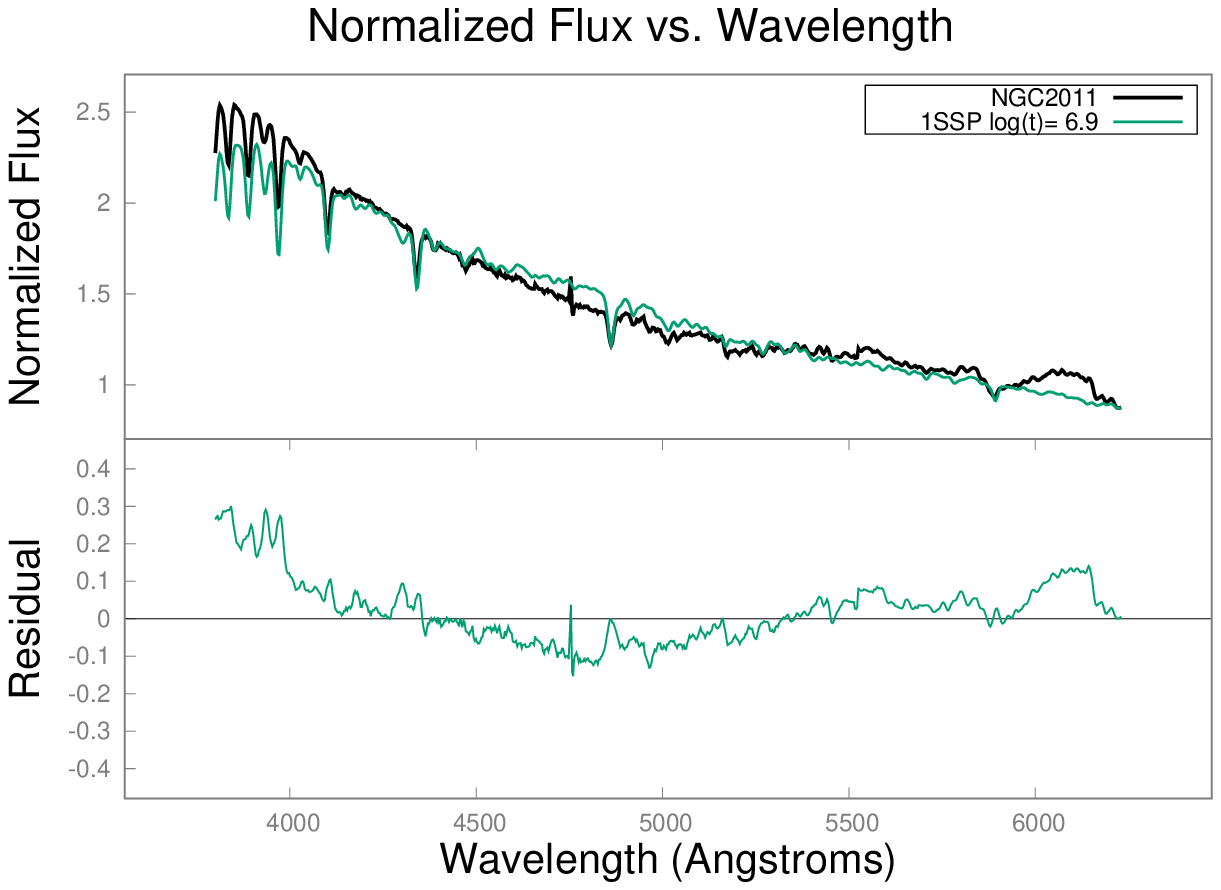}} \\
\end{tabular}
\caption{Best 1SSP fits achieved with MILES models (LMC metallicity and Kroupa Universal IMF) neglecting the extinction.}
\label{fig:SSP}
\end{figure}

To understand and validate our full spectrum fitting results obtained with single SSPs (1SSP),
using an extinction-independent analysis we 
compare the strengths of selected, prominent features measured in our clusters and in the models. In particular, to assess how significant are the residuals obtained for the molecular features we measure the TiO$_1$ Lick index at $\lambda\lambda \sim$5965~\AA. Figure~\ref{fig:indices} shows the index values after smoothing all the spectra to a common resolution of FWHM=14~\AA, which is similar to one of the resolutions proposed in the LIS-system \citep{Vazdekis10}. 
The TiO$_1$ index is maximized for ages around 10\,Myr, where the relative contribution from RSGs peaks, and the strengths of this molecular feature get larger for solar metallicity. The TiO$_1$ values of the clusters are significantly larger in comparison to the models (including those of solar metallicity in some cases), as already shown in the residuals obtained by our full spectrum-fits. This figure also shows that the high TiO$_1$ values obtained for the clusters cannot be matched by any combination of SSPs with LMC metallicity. 
It is worth noticing that, for reference, the TiO$_1$ index strength for our reference RSG star in the MILES database (HD\,42543; see Section~\ref{Sec_RSGs}) reaches a value as high as $\sim$0.2.

In the right panel of Fig.~\ref{fig:indices} we show the higher order Balmer line age indicator H$\delta_{A}$ (its wide index definition, which does not demand high signal-to-noise spectra). An advantage of the H$\delta$ feature with respect to the more standard age indicator H$\beta$ is that this index is much less affected by emission filling-in. 
According to the analysis performed in Section~\ref{Sec_ppxf} the spectrum of NGC\,1994 shows some emission contamination in the H$\delta$ feature (and stronger in H$\gamma$ and H$\beta$).  We do not find any significant emission in  NGC\,1984, whereas for NGC\,2011 this correction is virtually negligible. The comparison with the models shows ages below $\sim$15\,Myr, in good agreement with our full spectrum fits. It is worth noticing that H$\delta_{A}$ keeps raising up with increasing age until reaching a maximum at 200-500\,Myr, after which starts to drop down to reach similar values as those measured in our clusters for ages around 2\,Gyr (see \citealt{Vazdekis15}). We recall that none of these two indexes are affected by extinction according to the thorough analysis performed by \citet{MacArturh05}.

To summarize, our line-strength analysis shows that we are able to obtain age estimates in good agreement with the full spectrum fits. However the high TiO$_1$ index values measured in the clusters, which cannot be matched by our models with the LMC metallicity, suggest that there are problems with the modeling of the RSGs and with the Blue to Red Supergiants ratio. In Section~\ref{Sec_RSGs} we focus on the relative RSG contribution in more detail.

\subsubsection{Balmer lines emission filling-in: spectral fits with pPXF}
\label{Sec_ppxf}

Inspection of the balmer lines in the LMC cluster spectra show  emission fill-in
in NGC\,1994. This is most prominent in the H$\beta$ lines, and becomes less evident (as expected) in the higher-order balmer lines. To quantify the degree of emission in these spectra we used pPXF \citep{Cappellari04} in conjunction with GANDALF (Sarzi et al. 2006) to construct and fit a series of MILES model-based templates to the cluster data. The outputs of this exercise were the best-fit model templates for each cluster and the best-fit emission spectrum. Focusing on H$\delta$ (which is the balmer line least affected by emission in our spectra), we obtain a correction factor for emission (i.e., H$\delta$(true) / H$\delta$(observed)) of 1.41 for NGC~1994. The emission line spectrum of NGC~1994 once 
subtracted the continuum component is shown in Fig.~\ref{fig:ppxf}; the 
EW(H$\beta$) in emission of this galaxy is 5.86~\AA.  
In the case of NGC~2011 we do not find significant emission, so the emission seen in 
the residuals is probably an artifact of the fit between the models and the spectrum  which resembles an emission line.
We detect no emission in NGC\,1984. 
The present results are consistent with the analysis of \cite{Santos95} who found only an emission component in NGC~1994 with a similar EW.

The origin of such emission is unlikely to be due to recent SN explosions: the maximum Supernova rate (SNR) in our clusters is about $3 \times 10^{-4}$ SN/yr (see Sect.~\ref{Sec_sampling} for details about this estimate). However SN produce significant emission  during a short time after the explosion. Assuming  a 100 yr lifetime we have a mean number of 0.03 SN in the most optimistic case. The other possibility is a nebular contribution  due to the ionization of massive stars.  In this case, the  computations from \cite{CMH94} predicts a EW(H$\beta$) in emission between 6 and 5 \AA ~at ages between 7.8 and 8.2 Myr (i.e. $\log t \sim 6.9$). At such ages the nebular contribution to the continuum is about 5\%, decreasing to a 2\% at 10 Myr.

We conclude that that a nebular component (lines and continuum) is not present, or at least has no impact in our analysis for NGC~1984 and NGC~2011, and in principle these clusters would have ages larger than 8 Myr, which is compatible with our 1SSP fits. However, that NGC~1994 shows a nebular contamination suggests an age younger than 8 Myr, which is younger than the ones obtained by our fits. In addition, this cluster shows problems with the $E(B-V)$ estimates, and it is the one where the curvature in the residuals is the smallest. So we have decided to do not consider NGC~1994 in our study about the RSG and curvature of the residuals in the fits (although it is included in the UV and CMD analysis).

\section{Testing different scenarios to improve the fits}
\label{Sec_2SSP}

To understand the origin of the residuals obtained in section ~\ref{Sec_1SSP} here we perform a series of tests to attempt to optimize the matching of the model with the observations. In this section we fit for with multiple stellar populations and for different IMFs. 

Since we are interested in minimizing the residuals of the fits, 
we have chosen to consider no reddening correction in this part of the analysis. This applies to the present section and Sect.~\ref{Sec_RSGs}. This  allows us to reduce the parameter space in the fitting process. The residuals with and without reddening correction are similar. This is illustrated in Fig.~\ref{fig:SSP} which can be compared to Fig~\ref{fig:SSPred}; note that NGC 1994 had been excluded in this analysis as explained in Sect.~\ref{Sec_ppxf}. The ages obtained in this section should not be considered as valid cluster ages, but simply regarded as reference for the comparison in the various tests.

\subsection{Combinations including an SSP of 10 Myr}
\label{Sec_IMF_13}

\begin{figure}
\begin{tabular}{ccc}
\resizebox{75mm}{!}{\includegraphics[angle=0]{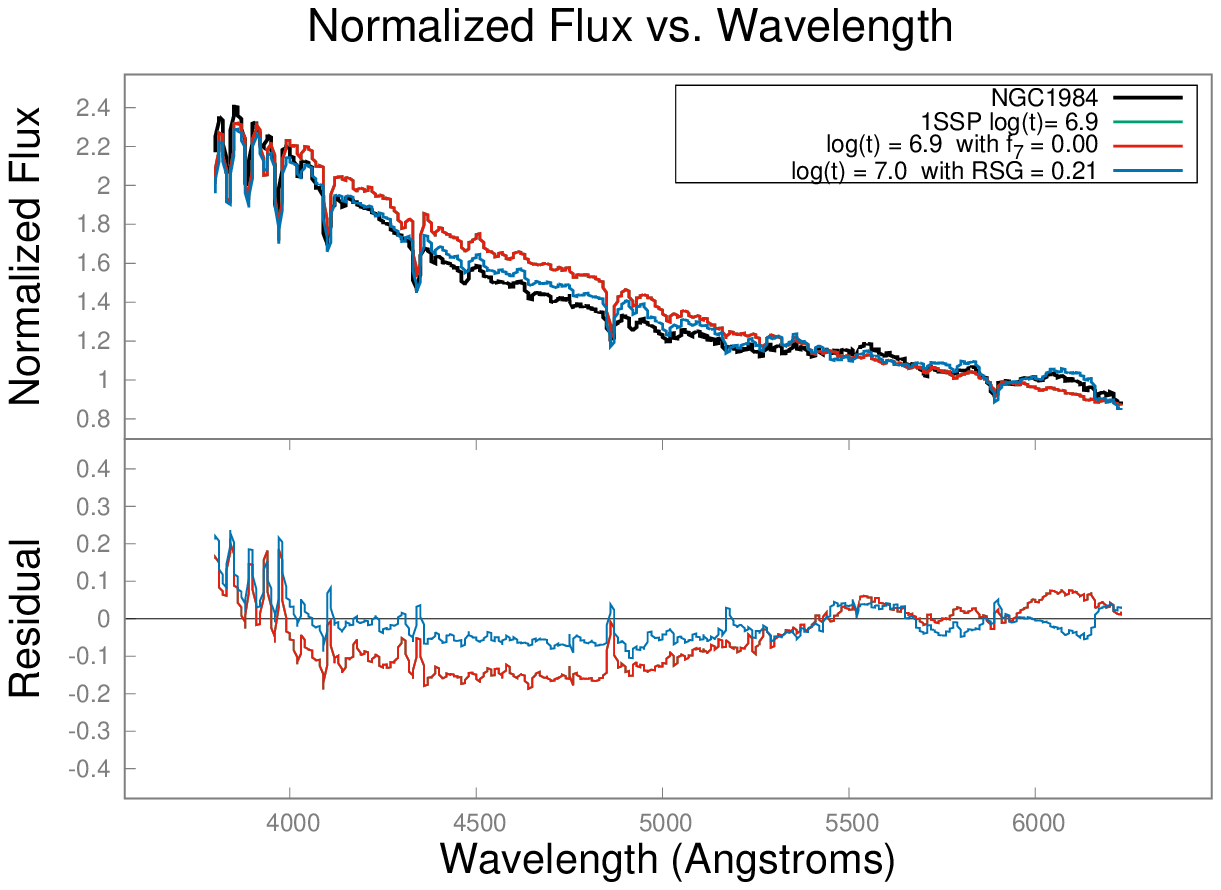}} \\
\resizebox{75mm}{!}{\includegraphics[angle=0]{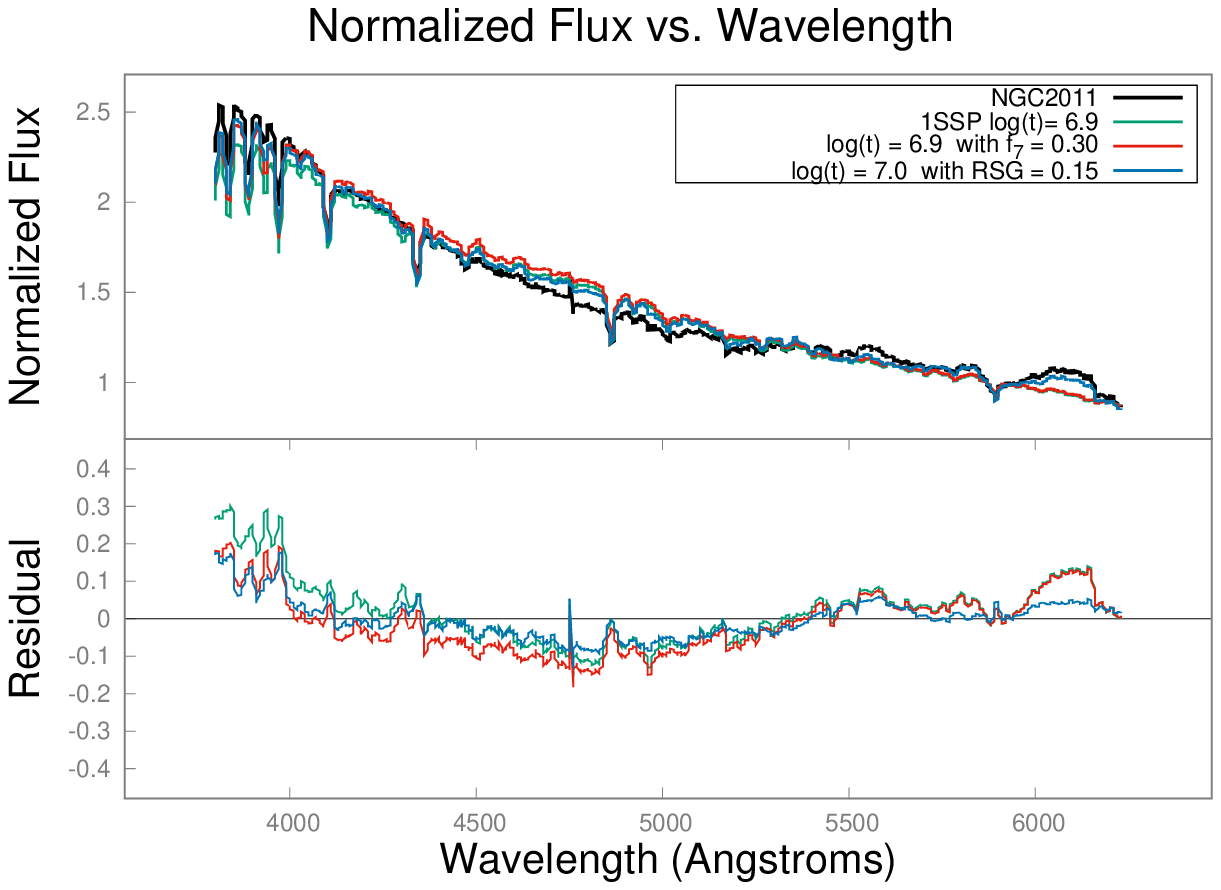}} \\
\end{tabular}
\caption{LMC metallicity results using MILES models and Kroupa Universal IMF neglecting extinction. The figure includes the reference result for a 1SSP fit in green, the the combination of two LMC metallicity SSP models where one of them has 10 Myr (c.f. Sect.~\ref{Sec_IMF_13}) in red, and the variation of the relative contribution of RSGs (c.f. Sect.~\ref{Sec_RSGs}) in blue.}
\label{fig:2SSP10MyrLMC}
\end{figure}

\begin{figure}
\begin{tabular}{ccc}
\resizebox{75mm}{!}{\includegraphics[angle=0]{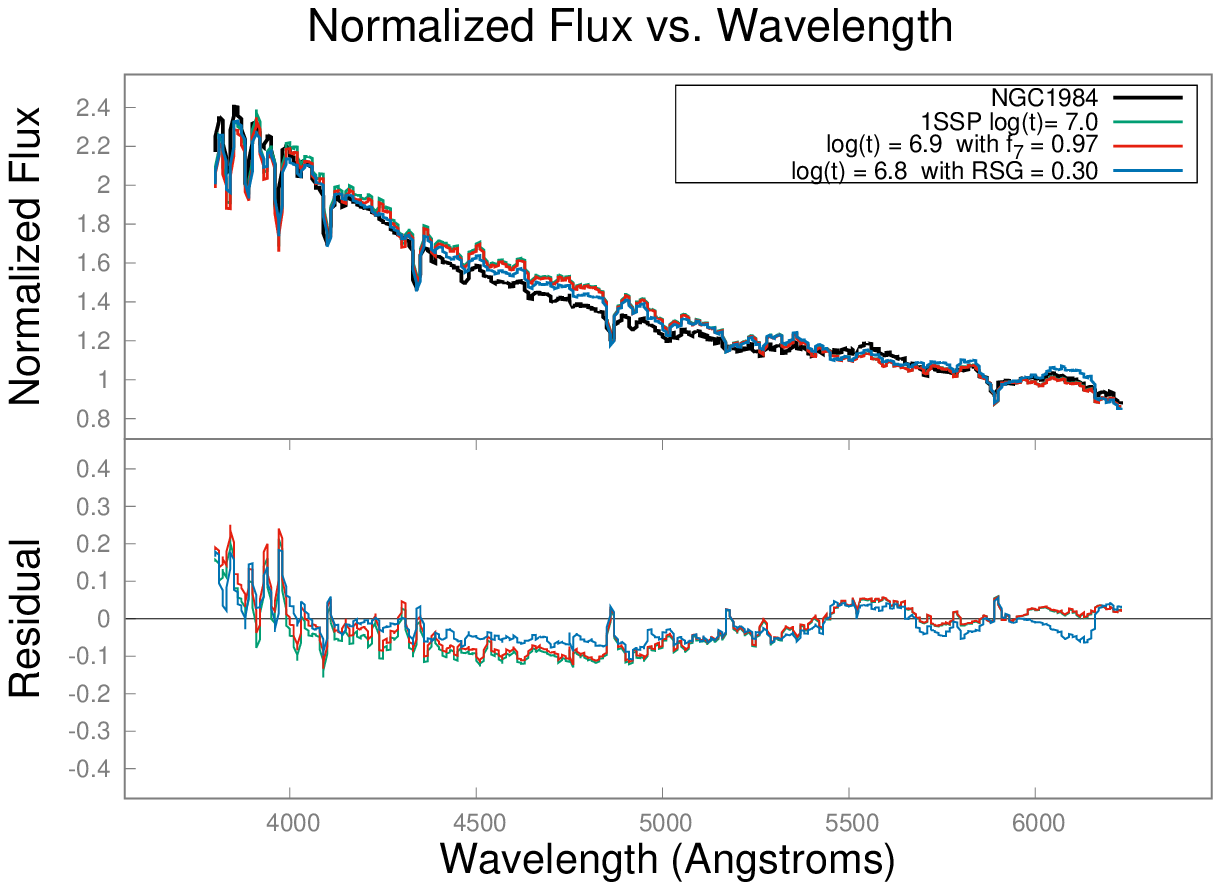}} \\
\resizebox{75mm}{!}{\includegraphics[angle=0]{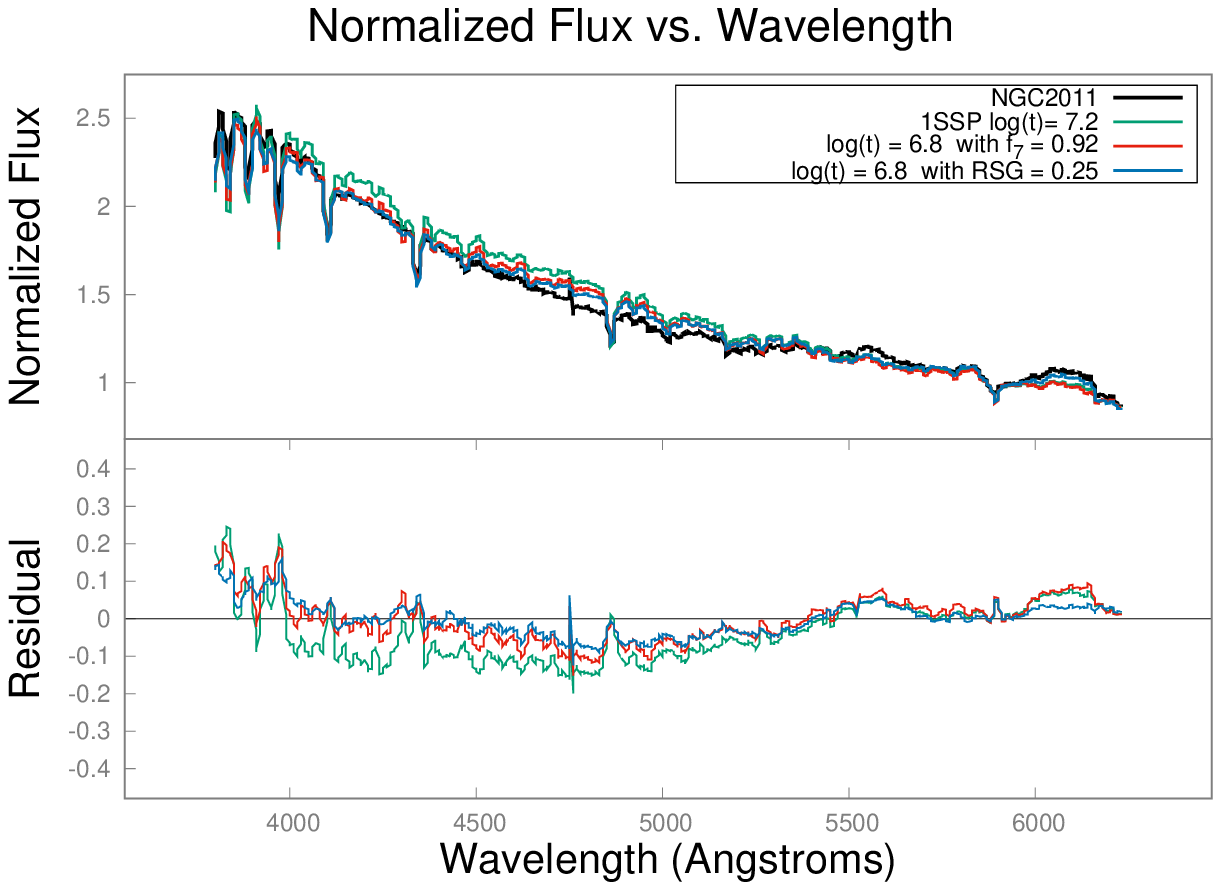}} \\
\end{tabular}
\caption{Solar metallicity results using MILES models and Kroupa Universal IMF neglecting extinction. The figure includes the reference result for a 1SSP fit in green, the the combination of two LMC metallicity SSP models where one of them have 10 Myr (c.f. Sect.~\ref{Sec_IMF_13}) in red, and the variation of the relative contribution of RSGs (c.f. Sect.~\ref{Sec_RSGs}) in blue.}
\label{fig:2SSP10MyrSOLAR}
\end{figure}

\begin{figure}
\includegraphics[angle=0,scale=0.45]{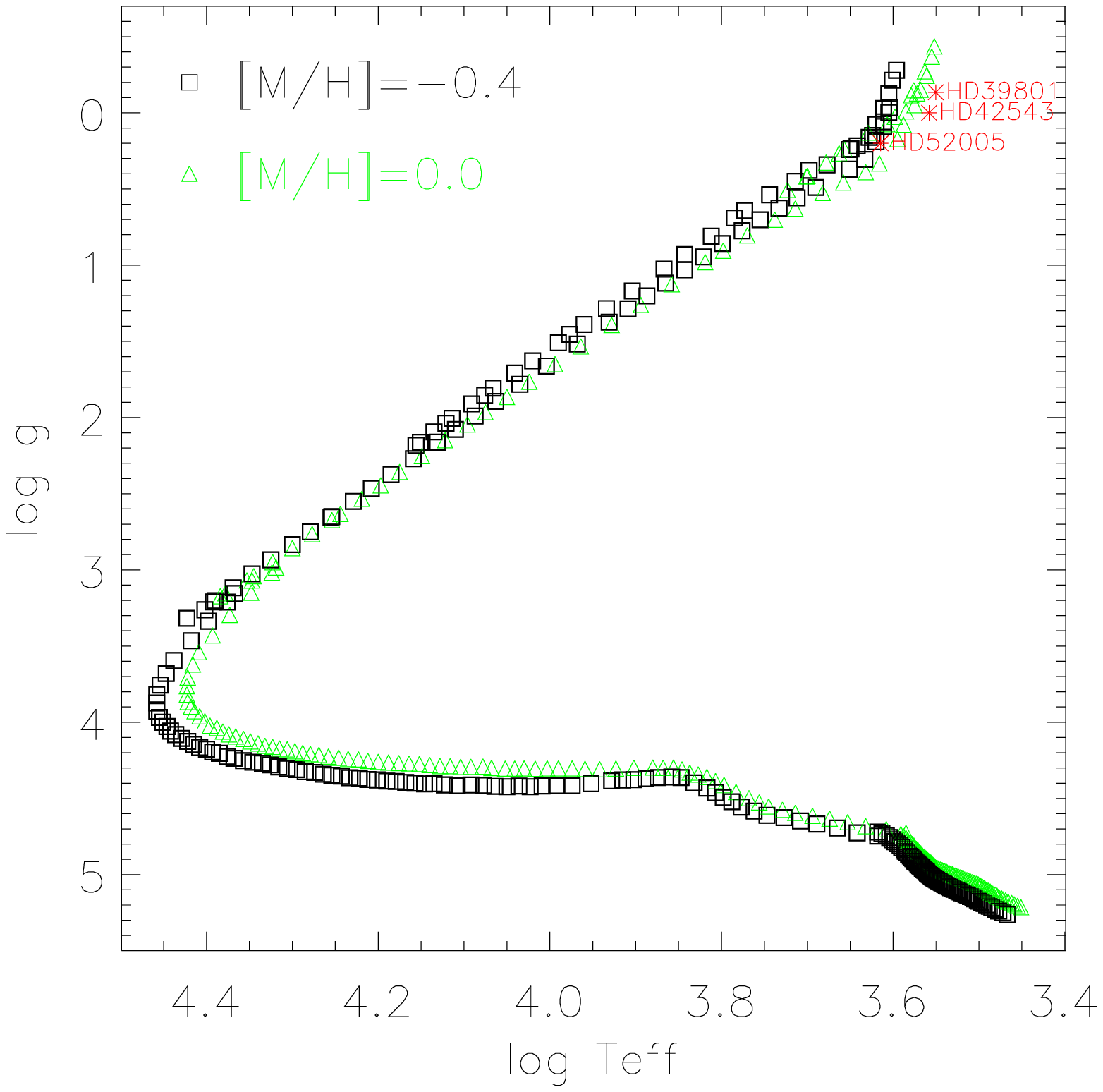}
\caption{Surface gravity - effective temperature plane showing isochrones of 10 Myr which feed our models (see the text for details).  Black squares draw an isochrone with metallicity [M/H]=-0.4 (Z=0.008, representative of the LMC), whereas the green triangles draw that of solar metallicity (Z=0.02). We also show three RSGs in the MILES stellar spectral library (Note that the spectrum of HD~42543 has been employed to perform the tests shown in
Section~\ref{Sec_RSGs}).}
\label{fig:isochroRSG.eps}
\end{figure}

A careful look at the SSP fits show the concavity in the continuum and the specific residuals related to the molecular features. This is  indicative of a significant contribution from red stars. In view of this, we constructed two-component modes where one component has an age of 10\,Myr. This age is chosen because at this age the relative contributions of the RSG stars to the total light is maximized in the SSPs with the LMC metallicity.
We used the MILES models with LMC
metallicity and Kroupa Universal IMF \citep {Kroupa01} to create spectral combinations according to:

\begin{equation} 
f_7*SSP(\log t = 7.0)+(1-f_7)*SSP(\log t)
\label{Eq_2SSPs}
\end{equation}  

\noindent where $f_7$ refers to the mass fraction contribution of an SSP with $\log t=7.0$. 
Both ages are kept fixed and only the value of $f_7$ is varied, running from 0 to 1 in steps of 0.01. Note that a change of 0.01 in mass fraction might
be relevant for these fits as it reflects a larger luminosity fraction,
depending on the age difference between the two SSPs. The results are given in
the last two columns of Table~\ref{tab:Results1} and is shown by the blue line in Fig.~\ref{fig:2SSP10MyrLMC}.
We include in this figure the best fits
achieved with a 1SSP (also neglecting reddening correction) for comparison, 
and a test about the variation in the contribution of RSGs performed in Sect.~\ref{Sec_RSGs}.
Figure~\ref{fig:2SSP10MyrLMC} shows that with these
combined models we are able to decrease the characteristic large wavelength
scale concavity of the obtained residuals for NGC~2011. These fits
also show that there is no significant improvement in the residuals seen in the
red spectral region, which are characteristic of strong molecular bands of the
RSGs. The latter might indicate that the RSGs contribution is not well reproduced
in the models. The quantitative results of the $\chi^2$ give minimum values for the RSG fitting. The green, red and blue lines give the $\chi^2$  values 7.73, 7.73 and 2.38 respectively for NGC1984 and 6.48, 4.95 and 2.16 respectively for NGC2011.

Judging from the residuals related to the molecular features in the red spectral
region that are obtained with the LMC metallicity with either one or two
components, the fraction of light that RSGs contribute to the total light in the models
seems to be underestimated (see also discusion in Sect. \ref{Sec_indices}). As standard stellar evolution
prescriptions predict an increasing contribution of red relative to blue
supergiants with increasing metallicity (e.g.,
\citep{Meylan82,CMH94,Langer95,Eggenberger02}) 
we also perform here fits with solar
metallicity (and Kroupa Universal IMF) MILES models to see if we obtain a better
match to our cluster spectra. We therefore repeat the analysis applying
Eq.\ref{Eq_2SSPs} but replace the LMC metallicity with solar metallicity. The
results are given in the last three lines of Table~\ref{tab:Results1} and Fig.~\ref{fig:2SSP10MyrSOLAR}. The new fits
indicate a slight decrease of the concavity residuals and a significant
improvement in the molecular features related residuals. The quantitative results of the $\chi^2$ give minimum values for the RSG fitting. The green, red and blue lines give the $\chi^2$  values 3.66, 3.51 and 1.99 respectively for NGC1984 and 5.86, 3.00 and 1.64 respectively for NGC2011. Such improvement is not
necessarily due to the change in metallicity itself, but is most likely due
to the change in the BSG to RSG ratio associated with this metallicity increase.
This is illustrated in Fig.~\ref{fig:isochroRSG.eps}, where we show the theoretical isochrones (see Section~\ref{Sec_models}) corresponding to 10\,Myr for both the LMC and Solar metallicities.  Note that the RSGs of the solar metallicity isochrone reach significantly cooler temperatures as shown in the upper right corner of the figure. We also include in the plot a set of representative MILES RSG stars. In particular, the star HD\,42543, which has a very similar metallicity to that of the LMC (see \citealt{Cenarro07} and references therein), is cooler than predicted by the isochrone with the LMC metallicity. We discuss the effects of varying specifically the RSGs in Section~\ref{Sec_RSGs}.
  
\subsection{Fitting with a top-heavy IMF}
 \label{Sec_IMF_08}

\begin{table*}
\caption{Results Using $\Gamma_b$ = 0.8}
 \label{tab:Results2}
 \begin{tabular}{lccccccc}
 \hline
Name & $\Gamma_b$ & Metallicity & Age$^1$ & E(B-V)$^1$ & Age$^2$ & Age$^3$ & $f_7$\\
 &  &  & $\log(\mathrm{age/yr})$ &  & $\log(\mathrm{age/yr})$ & $\log(\mathrm{age/yr})$ & \\
\hline
NGC~1984 & 0.8 & LMC & 6.9 & 0.00 & 6.9 &  6.9  & 0.19 \\
NGC~2011 & 0.8 & LMC & 7.0 & 0.07 & 7.5 &  6.9  & 0.66 \\
NGC~1984 & 0.8 & Solar & 7.2 & 0.03 & 7.0 & 6.8 & 0.94 \\
NGC~2011 & 0.8 & Solar & 7.3 & 0.03 & 7.2 & 6.8 &  0.84 \\
NGC~1984 & 0.3 & Solar & 7.2 & 0.00 & 7.2 &  6.8 & 0.89 \\
NGC~2011 & 0.3 & Solar & 7.3 & 0.00 & 7.3 &  6.8 &  0.80 \\
\hline
\multicolumn{8}{l}{$^1$ Using 1SSP with reddening correction.}\\
\multicolumn{8}{l}{$^2$ Using 1SSP neglecting reddening correction.}\\
\multicolumn{8}{l}{$^3$ Using equation \ref{Eq_2SSPs}}
\end{tabular}
\end{table*}

\begin{figure}
\begin{tabular}{ccc}
\resizebox{75mm}{!}{\includegraphics[angle=0]{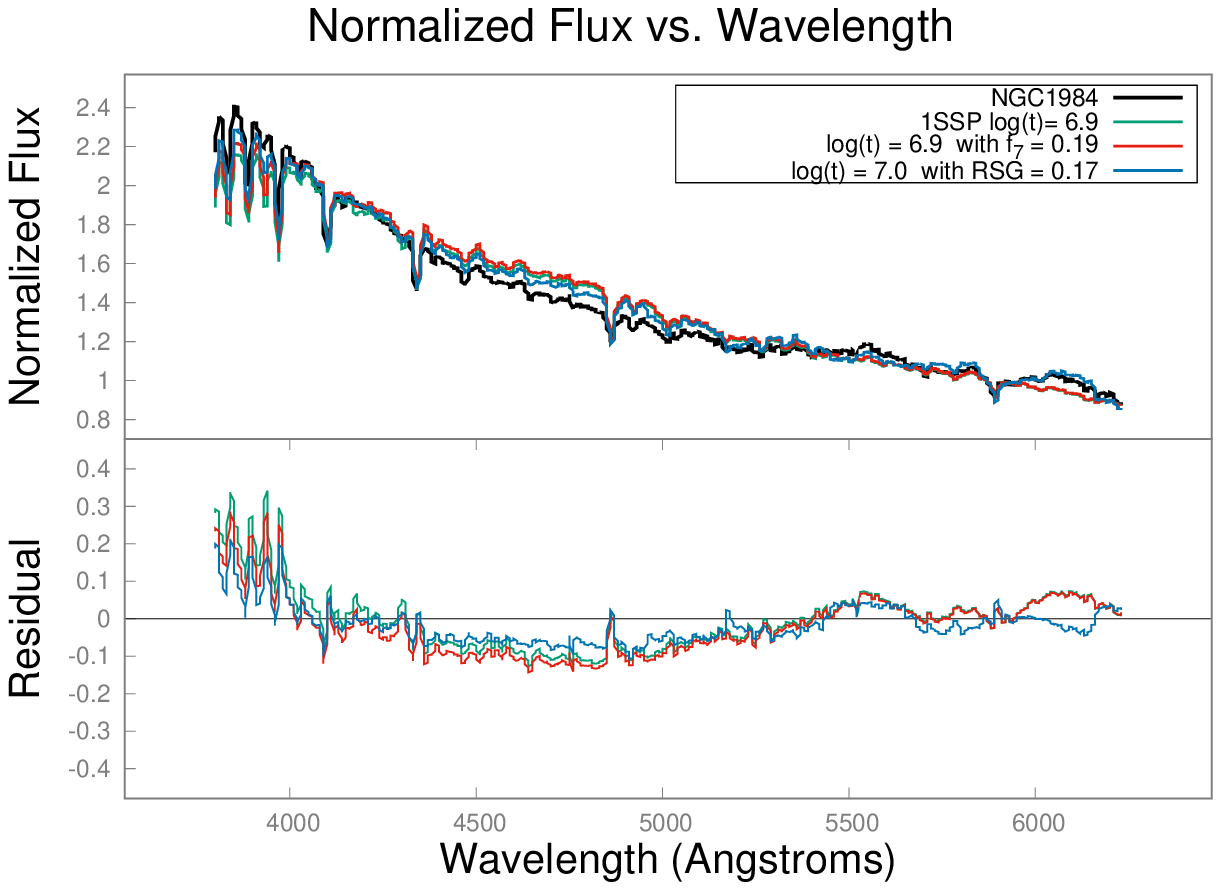}} \\
\resizebox{75mm}{!}{\includegraphics[angle=0]{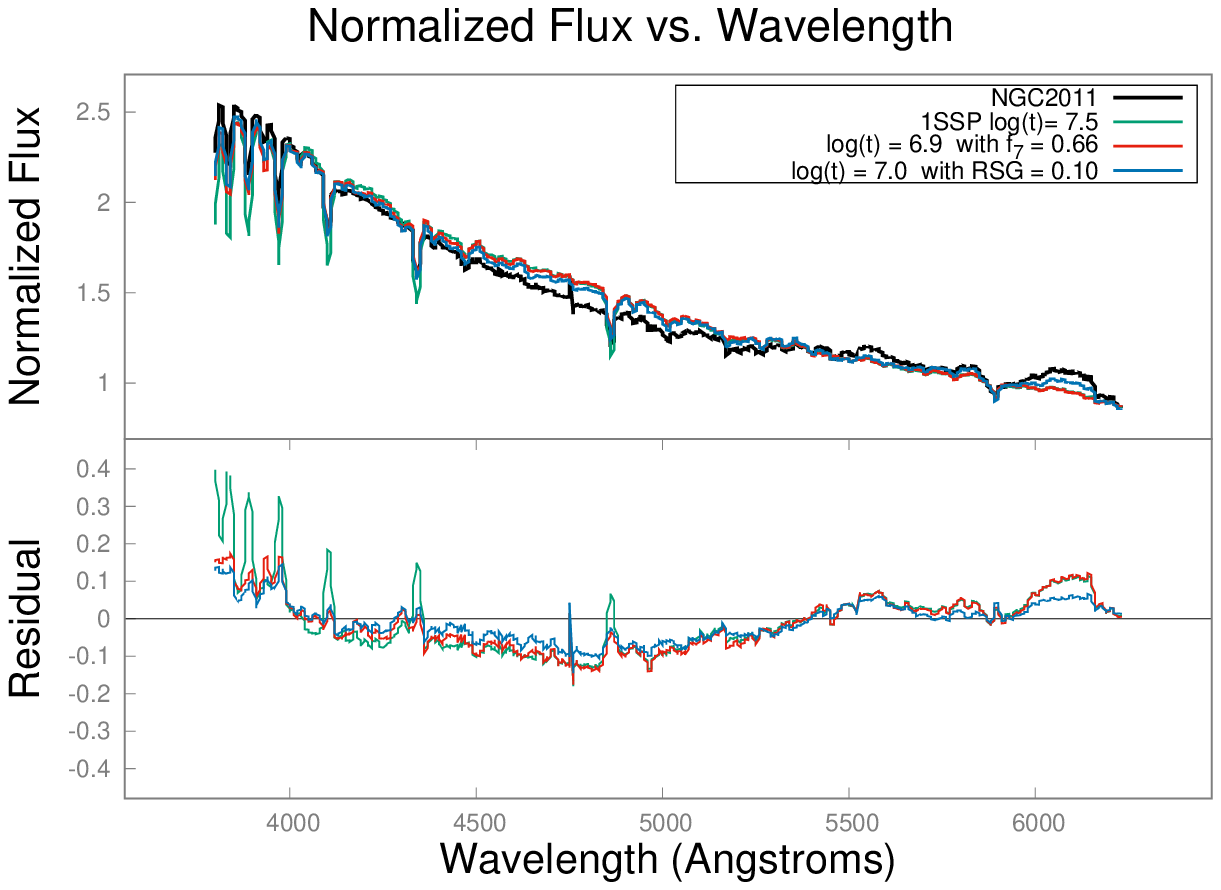}} \\
\end{tabular}
\caption{The results obtained using MILES models with LMC metallicity and IMF = 0.8 as described in Sect.~\ref{Sec_IMF_08}.}
\label{fig:43}
\end{figure}

We also tested whether varying the IMF could improve our fits. We choose a flatter
IMF shape than Kroupa, i.e. top-heavy, as such an IMF will increase the relative number
of massive, more evolved, stars. In particular we tested the match of the SSP
with the clusters using a Bimodal IMF with slope $\Gamma_b$ = 0.8 with LMC
metallicity first, then with solar metallicity. The results are shown in
Fig.~\ref{fig:43} where the quantitative results of the $\chi^2$ give minimum values for the RSG fitting. The green, red and blue lines give the $\chi^2$  values 5.84, 5.36 and 2.31 respectively for NGC1984 and 6.63, 4.15 and 2.17 respectively for NGC2011.
The best matching models parameters are given in Table~\ref{tab:Results2}. 
No significant improvements are seen in comparison to the equivalent fits obtained with a Kroupa Universal IMF. 

We performed fits with an even flatter bimodal IMF with $\Gamma_b$ = 0.3 and solar metallicity.
The fits show slightly worse agreement with respect to those with $\Gamma_b$ = 0.8.  
We conclude that IMF variations cannot explain the spectral characteristics of
these clusters.

\section{Varying the relative contribution of RSGs}
\label{Sec_RSGs}

In this Section we aim to provide quantitative estimates of the RSG
contribution as derived from our fits to the data. For this purpose we use the
spectrum of a representative RSG from the MILES library with similar metallicity
to that of the LMC. The selected star is HD~42543 shown in
Fig.~\ref{fig:isochroRSG.eps} with other RSGs that are also present in the MILES
library. Note that this star has a cooler temperature than predicted by the
isochrone with the LMC metallicity. We therefore combine the SSP spectrum with
HD42543 according to the following equation:

\begin{equation} 
f_{RSG}*(RSG)+(1-f_{RSG})*SSP(\log t)
\end{equation}  

\noindent where $f_{RSG}$ refers to the additional contribution in light from
RSGs that is required by the SSP. $f_{RSG}$ varies from 0 to 1 in steps of 0.01.
The results we obtain indicate a log (age/year) of 7 for both clusters with a luminosity fraction of 21\% and 15\% for NGC~1984 and NGC~2011 respectively. This is shown in Fig.~\ref{fig:2SSP10MyrLMC} and Fig.~\ref{fig:2SSP10MyrSOLAR}.

We notice that the RSG combinations improve the fitting of the clusters over
those achieved with either one or two SSPs. Comparing the residuals of Fig.~\ref{fig:2SSP10MyrLMC} and Fig.~\ref{fig:2SSP10MyrSOLAR} we see that both the residuals of the large scale
concavity and the molecular features in the red spectral range are significantly
smaller. Moreover, the residuals shown in Fig.~\ref{fig:2SSP10MyrLMC} and Fig.~\ref{fig:2SSP10MyrSOLAR}
are  similar (although with different contributions from the additional
RSGs) irrespective of the metallicity of the SSP, which highlights the
role of the RSGs. These results explain why the fits with one SSP or even with
two SSPs improve significantly only for solar metallicity, as the relative
contribution of RSGs increases.
These fits show that the most likely source for the mismatch obtained when
fitting with standard SSPs is the fact that at sub-solar metallicities, such
as that of the LMC, the relative contribution from RSGs is not properly predicted by the
stellar evolution models (either Padova or Geneva) implemented in the SSP
models. Although not tested here we also note that this also applies to models
with stellar rotation \cite[e.g.,][]{Maeder00}, such as those recently employed
in the stellar population models of \citet{Leitherer14}. These models, which are
also based on the Geneva tracks evolutionary prescriptions, do not provide
significantly different blue to red supergiants ratios in the LMC metallicity
regime (see \citealt{Meynet13} for more details) and, therefore, the results
would not differ significantly from those 
obtained using Geneva based Granada models,
except for the obtained age of the best matching SSP. Stellar rotation will
increase the lifetime of the stars and therefore the net contribution of the RSG
stars will peak at older ages, a few Myr beyond the $\sim$10\,Myr, with respect to
the non-rotating, standard, case.
To summarize, the best fits obtained with the LMC metallicity
require an increase in the RSG contribution of around 20\% with respect to the
prescription of the SSPs for NGC~1984 and NGC~2011.

\subsection{Varying the RSG contribution using partial SSPs}
 \label{Sec_pSSPs_RSGs}

  \begin{figure}
\begin{tabular}{ccc}
\resizebox{75mm}{!}{\includegraphics[angle=0]{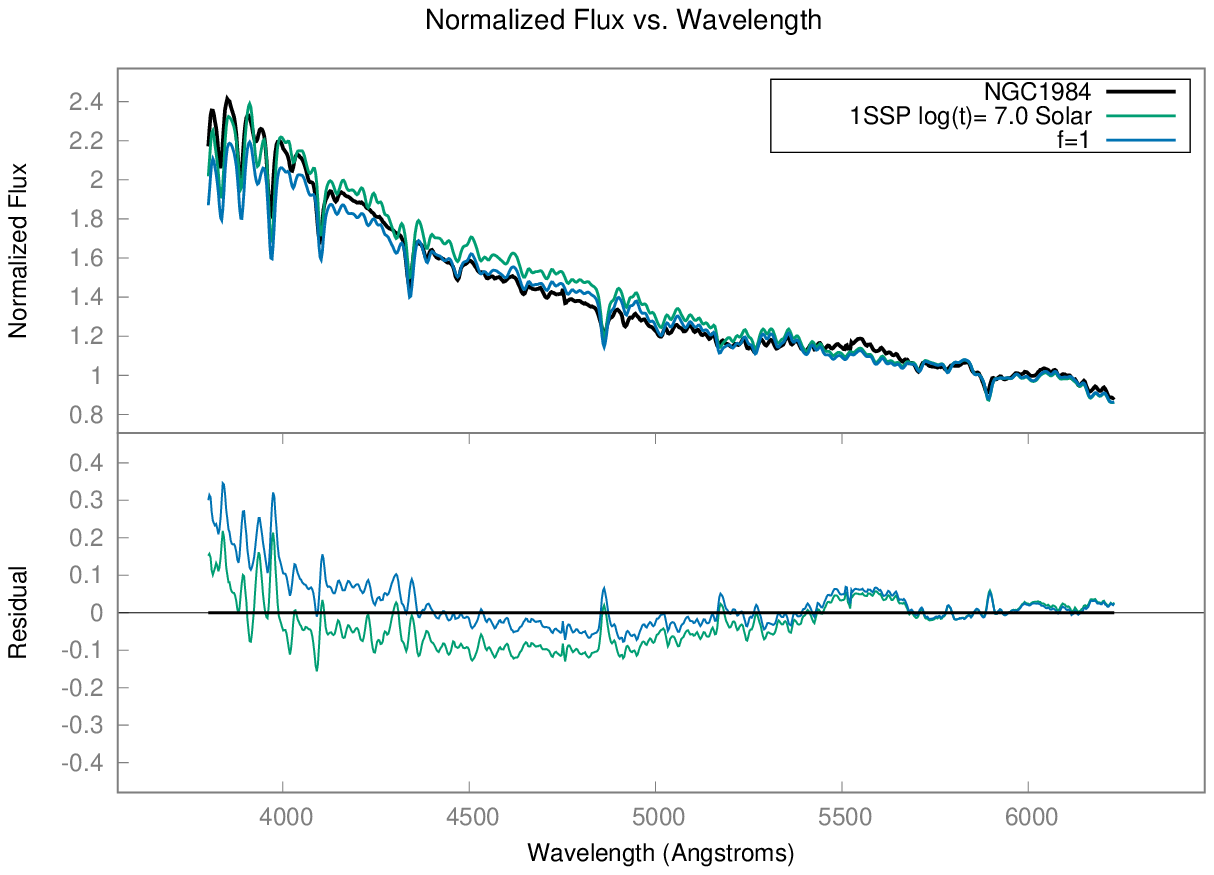}} \\ 
\resizebox{75mm}{!}{\includegraphics[angle=0]{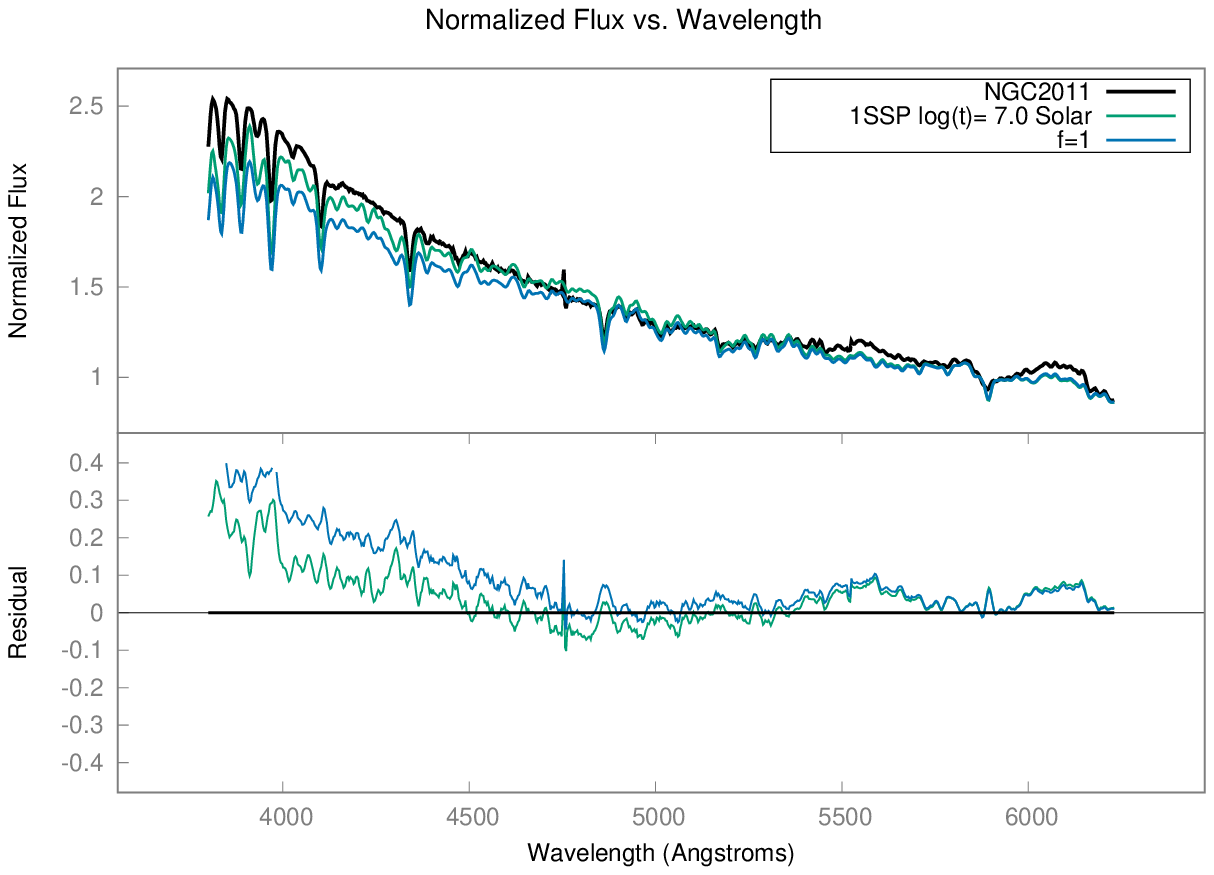}} \\
\end{tabular}
\caption{The results obtained using equation \ref{Eq_pSSPs_RSGs} with partial SSPs for masses below/above 17\,M$_{\odot}$ where the stars below 17 M$_{\odot}$ have evoleved with LMC metalliciy and stars above 17 M$_{\odot}$ have evolve with solar metallicity as explained in Sect.~\ref{Sec_pSSPs_RSGs}.}
\label{fig:5}
\end{figure}

Here we use partial SSPs to test whether by substituting the prescription of the
latest evolutionary stages for the LMC metallicity with that of the solar
metallicity we improve the fits to our clusters. For this purpose we computed
pSSPs below and above 17\,M$_{\odot}$. For 10\,Myr the MS turnoff takes place
around 15\,M$_{\odot}$, whereas the reddest RSGs have around 20\,M$_{\odot}$.
Therefore the selected mass corresponds roughly to the evolutionary stage
located half the way in between these two stages. 

We therefore hypothesize that our modeling for the more evolved
stars, i.e. pSSP$_h$ might be improved if using that corresponding to the solar
metallicity. We test the combination expressed in the equation below by varying
the relative contribution, $f$, of the more evolved stars that include the RSGs

\begin{equation} 
pSSP_{l}(10Myr,LMC) +  f*pSSP_{h}(10Myr,Solar)
\label{Eq_pSSPs_RSGs}
\end{equation} 

\noindent with this fraction varying in between 1 and 2. This choice is
motivated by the fact that our fits with the SSPs (and specifically with the RSG
star spectrum) have shown us that we need a higher contribution from the RSGs.
In practice this excercise means roughly that we exchange the more evolved phase
of the isochrone with LMC metallicity with that of the solar metallicity (see
Fig.~\ref{fig:isochroRSG.eps}). The results are shown in Fig.~\ref{fig:5}. The best fitting gives an f value of 1.

We also performed a more extreme test by computing pSSPs below and above
8\,M$_{\odot}$. This mass represents roughly the transition from intermediate to
high mass stars. Therefore in this test we consider the possibility that our
modelling of all massive stars might not be appropriate. 
Our analysis shows that we can remove to a great
extent both the concavity and the molecular band residuals although we do not
match the bluest spectral region. This result also confirms those shown in
Section~\ref{Sec_RSGs}, that we need to enhance the relative contribution of
RSGs for the LMC metallicity. 

\subsection{Raising the temperature of the RSGs}

To close this section, we also performed tests using a set of MILES models in which we artificially raised the effective temperature for all stars with $\log g < 0.6$ and $T_\mathrm{eff} <$ 4000K by 200K in the MILES library. For the RSGs with effective temperature between 4000K and 4500K we raised the $T_\mathrm{eff}$ by 100K.  This allows us to test the possibility suggested by \citep {Lee15, Lee16} that the actual effective temperature estimates for the RSGs is cooler by 200K. 
Our fits show that increasing the $T_\mathrm{eff}$ of the RSGs it is not enough to decrease the residuals significantly with respect to adopting the standard $T_\mathrm{eff}$. 

\section{Cross-checking with other approaches}
\label{Sec_checking}

In this section we made an independent cross-check of the cluster parameters (age, extinction and mass) using a UV and CMD analysis. We note that in the UV spectral range our analysis is insensitive to the RSGs, unlike in the study
based on the optical range performed in Sections 4,5 and 6. Moreover, the CMD analysis mainly
deals with the fit of the main-sequence and main sequence turn-off, hence is not affected by post main sequence evolutionary phases.  These two independent studies that we perform here will therefore help us at constraining our
full spectrum fitting (and line-strength) solutions obtained from the optical range.

\subsection{UV independent analysis: mass, extinction and sampling issues}
\label{Sec_sampling}

\begin{figure}
  \resizebox{75mm}{!}{\includegraphics[]{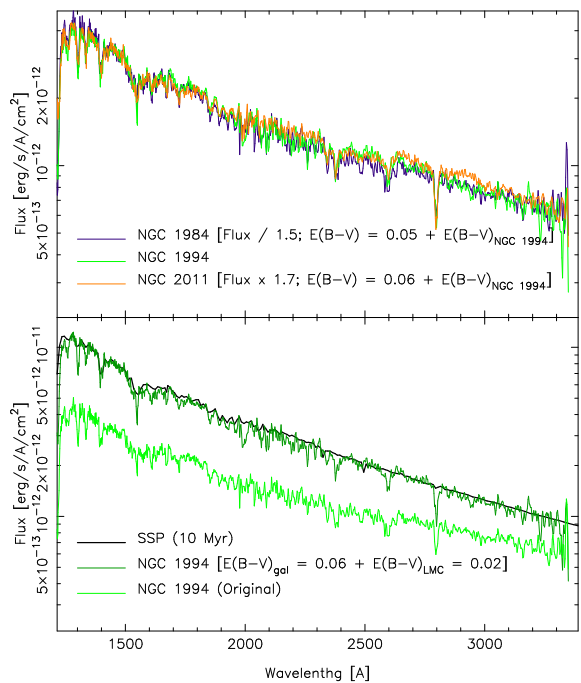}}
\caption[]{{\it Top panel:} Comparison of the three clusters to provide a similar UV spectra by variations of $E(B-V)$ values among them. {\it Lower panel:} $E(B-V)$ estimate for the cluster to be compatible with an SSP of 10Myr.}
\label{fig:UV_clus}
\end{figure}

We have shown that each of our clusters has its own peculiarities which can be attributed to varying RSG
contributions, even for the case where the clusters have a similar age. In principle this
might be reflecting that here are not enough stars to sample the post main sequence
of the cluster, and that our analysis, which assumes that there are no sampling problems, could be affected.

As shown by several authors (\citealt{CL04}, \citealt{Cervin06}, \citealt{PH10} among 
others, see  \citealt{Cer13} for a review on the subject)
sampling effects would fool the analysis using standard methods. 
In a brief summary, the integrated luminosity described by an SSP is actually a probability distribution of possible integrated luminosities, being the standard SSP result the mean value of such distribution, whose shape varies with the observed number of stars (hence total mass). The validity of $\chi^2$ fitting methods depends on the shape of this distribution. 
An extreme case when $\chi^2$ must be ruled out is when the model includes stars more luminous than the cluster itself \cite[the Lowest Luminosity Limit test presented by][hereafter LLL]{CL04}, and the effects are proportionally reduced as the observations contains a larger number of stars.
In addition, such sampling effects are wavelength dependent, since the  the number of stars which contribute effectively to a given wavelength varies with wavelength and age \citep{Buzz93,Cervin02,Cer13}

For our particular case stochastic effects are lower 
 in the UV than in the optical \cite[see fig. 11 and discussion in][ which is a valid guide line for our 
 case although obtained for solar metallicity]{Cer13}.  As our optical spectra are not flux 
 calibrated in absolute units, we cannot obtain mass estimates from the data. Since the IUE 
 aperture is similar to one used in the optical data, we use the UV spectra to give a mass estimate that can also apply to the optical. The present analysis is based on the by-eye study of the  overall IUE spectral shape  of the clusters in the 1250 to 3000 \AA ~wavelength range assuming a SSP, since we only aim to quote the range of UV compatible solutions with the other approaches discussed in this work.
We recall that the UV analysis has been performed with  {\sc{sed}@} synthesis code  \cite[][and references therein]{MHK91,Cervin06} using the same IMF and isochrones as in the MILES models, but with the low resolution BaSeL2 theoretical atmosphere  library from \cite{Lejeune97,Lejeune98} with LMC metallicity. Such code provides, beside the SED, the relevant statistical quantities, as the LLL, ${\cal N}_\mathrm{eff}$, skewness and kurtosis, used in this section.

\begin{table}
\caption{Cluster properties compatible with a common age of 10 Myr for all clusters obtained from the UV analysis.}
\label{tab:UV10Myr_results}
\begin{tabular}{lccc}
\hline
 & NGC 1984  &  NGC 9194 & NGC 2011\\ 
\hline
Max. $\log t^1$  &  7.9  & 7.7  & 8.0   \\
Max. $\log t^2$  & 7.6 & 7.3 & 7.8  \\
$E(B-V)_\mathrm{GAL}$ & 0.06 & 0.06 & 0.06 \\
$E(B-V)_\mathrm{LMC}$ & 0.07 & 0.02 & 0.08 \\
$E(B-V)_\mathrm{tot}$     & 0.13 & 0.08  & 0.14 \\
$M$   [M$_\odot$]             & $7.5 \times 10^5$ & $5 \times 0^5$ & $3 \times 10^5$ \\
\# $M > 8 \mathrm{M}_\odot$      &    6382      &  4255       &   2553     \\
\# $M > 17 \mathrm{M}_\odot$      &  780        &     520    &    312    \\
\# Post-MS                              &    170     &   113   &    68    \\
\# BSG                              &      84     &     56   &    34     \\
\# YSG                              &      26     &     17   &     10   \\
\# RSG                              &      59     &     40   &     24   \\
${\cal N}_\mathrm{eff}$(1365\AA)     & 2184   & 1456   &   874   \\
${\cal N}_\mathrm{eff}$(4000\AA)     &      59   &   39    &     24   \\
${\cal N}_\mathrm{eff}$(6000\AA)     &      34   &   22    &     14  \\
$\left< SNr \right>$ [SN/yr]     &      $3.2\times 10^{-4}$   & $2.1\times 10^{-4}$     &  $1.3\times 10^{-4}$   \\
\hline 
\multicolumn{4}{l}{$^1$ Age estimated using the UV shape without constraints}\\
\multicolumn{4}{l}{$^2$ Age estimated using the UV shape constrained to $E(B-V)_\mathrm{GAL}= 0.06$}
\end{tabular}
\end{table}

\subsubsection{Ages and sampling effects estimates from UV data without constraints}

Mass estimates are obtained comparing the model results with the un-reddened observed flux and therefore depend on the assumed age and extinction values. 
Adopting the extreme case of unknown ages, we compared the LLL of the youngest age of the models (3.98 Myr) with the observed IUE data taking into account the distance to the LMC. We use the flux level at 1365 \AA ~as reference value of the UV since it maps the contribution of most massive stars in the system in a region without strong spectral features, hence a good proxy of the continuum level 
\cite[c.f][]{MHK91,MHK99}. For our clusters, NGC 2011 is 10 times above the LLL, and NGC 1984 and NGC 1994 are 30 times above the LLL; hence, at least for the UV analysis, there are no  strong sampling effects. 

For any older age, the situation improves, since (a) the level of the most 
luminous star which defines the LLL, decreases with age (b) the reference UV flux obtained by synthesis models decreases with time, and (c) there is not a strong variation of the sampling properties in the considered age range. This results in a larger inferred mass and a lower impact of sampling effects given that the observed flux has a fixed value.

Neglecting sampling effects for the moment, the shape the SSP UV flattens monotonically  with increasing age and extinction. Therefore we can establish an upper age limit for the cluster
by comparing the observed shape without any extinction correction with the SSP shape at different ages:  the limiting age is the first one where, at wavelengths larger that 1365 \AA (where observational an theoretical SED have a common value), the SSP UV SED is above the observed one;  such limiting age is typically defined by the SED at wavelengths larger than 2500 \AA. As results we find an upper limit of the age of $\log t = 7.9$, 7.7, and 8 for NGC~1984,  NGC~1994, and NGC~2011 respectively. If we take into account the value of $E(B-V)_\mathrm{G}$ in the direction of the LMC, the upper value for the ages are $\log t = 7.6$, 7.3, and 7.8 for NGC~1984,  NGC~1994, and NGC~2011 respectively. The results are show in the two first rows in Table \ref{tab:UV10Myr_results}.

\subsubsection{Mass and extinction estimates from UV data for a 10 Myr SSP}

Now we determine the cluster masses, which are age and extinction dependent.
To simplify the process we first compare the UV spectra of the three sources 
in order to obtain a single reference UV spectra with a common continuum shape.
To do this, we corrected all the observed spectra for galactic extinction and we varied the $E(B-V)_\mathrm{LMC}$ extinction using the \cite{Fit85} extinction law. We have used the spectra of NGC~1994 as reference since it is the steepest one. The results are shown in the top panel of Fig. \ref{fig:UV_clus}: the UV slope of the three clusters are compatible each other (which small discrepancies for NGC 2011 in the 2600 to 3000 \AA ~region) just by using $E(B-V)_\mathrm{LMC}$ variations. More interestingly, the three spectra have similar absorption features for Si~{\sc iv} 1400 \AA, C~{\sc iv} 1550 \AA ~and Mg~{\sc i} 1800 \AA. 
This provides some information about the stellar
populations present there:
The Mg~{\sc i} 2800 \AA ~absorption line is an indicator of the presence of blue supergiants \citep{Fanelli92}. 
Unfortunately we cannot obtain much more information from this line since a gap in the atmosphere models library at $\log g < 3.5$ at $\log T_\mathrm{eff} > 4$ which corresponds to the B and A supergiant region, so the line is not reproduced by the SSP modeling at such ages. However, the Si~{\sc iv} 1400 \AA, and  C~{\sc iv} 1550 \AA ~features allows us to establish some age ranges (assuming no strong sampling effects):
These features do not show evidence of the presence of P-Cygni profiles, hence very massive hot young stars are not present in the cluster, which implies ages larger than 5 Myr ($\log t=6.74$). In addition the ages cannot be much older than 40 Myr ($\log t= 7.6$) since both features disappear with age.

We stress that we are not doing an exhaustive analysis of the UV, neither that such results imply that a common age for the three clusters can be assumed (we know that this not the case). We simply take advantage of the smooth and monotonic variation of the UV spectra at ages larger than 5 Myr. A similar experiment using optical data shows that the spectral shape of the three clusters cannot be reduced to a common spectra just by $E(B-V)$ variations.

With the reference observed spectra, we can obtain inferences that can be translated to the specific clusters including mass inferences: NGC~1984 is 1.5 times more massive with an extra extinction of $\Delta E(B-V)=0.05$ and NGC~2011 is 1.7 times less massive with an extra extinction of $\Delta E(B-V)=0.06$ with respect the reference spectra, which is the one of NGC~1994. We show in Fig.~\ref{fig:UV_clus} the comparison with a 10 Myr SSP if it has an extinction of $E(B-V)_{\mathrm{NGC}\,1994} = 0.08$, (where $E(B-V)_\mathrm{G} = 0.04$ and $E(B-V)_\mathrm{LMC} = 0.02$).

With these numbers in hand and taking into account the distance to the LMC and assuming a 10 Myr SSP model, we can estimate cluster masses 
and the absolute number of stars of different types, as well as the effective number of stars which contributes to
1365, 4000, and 6000 \AA. We note that due our methodology fitting with steps of $\Delta E(B-V) = 0.01$, our mass estimates have an error of at least 25\%. This error increases with age variations, but since lower ages also imply larger $E(B-V)$ values there is no dramatic change in our estimates, and they do not reach a variation larger than a factor of 2 in any case.

Results for 10 Myr are shown in Table \ref{tab:UV10Myr_results}, where we have defined as Post-MS stars all those stars
with $\log g < 3$; from this group,
we define BSG those with $\log T_\mathrm{eff} > 4$, YSG those with $\log T_\mathrm{eff} \in [3.628,4]$ where 
$\log T_\mathrm{eff} = 3.628$ correspond to $T_\mathrm{eff} = 4250 \mathrm{K}$, and RSG as stars with 
$\log T_\mathrm{eff} <  3.628$. We also quote in Table \ref{tab:UV10Myr_results} the stars with masses above 
17 and 8 M$_\odot$ used in our pSSP analysis in Sect. \ref{Sec_TestpSSP}.  We also  compute the average  Supernova rate from the last point in the isochrones at $\log t=6.9$ and  $\log t=7$ in the last row; this value has been discussed in Sect.~\ref{Sec_ppxf}.

With these masses, the distribution function of integrated luminosities of the 10 Myr SSP in the optical wavelength has
skewness about $<0.1$ and a kurtosis about $<0.01$ in the worst case whereas the values at 1365\AA are 3 and 15 times 
lower respectively. For such values
the mean and mode values are almost similar so our $\chi^2$ analysis is not compromised.
In addition, the values of ${\cal N}_\mathrm{eff}$ at different wavelengths are consistent with 
with our estimates of stellar numbers, however, we caution that a direct comparison of both quantities 
cannot be done:  ${\cal N}_\mathrm{eff}$ is just an statistical measure of the dispersion
of the integrated luminosities at a given wavelength ($1/\sqrt{{\cal N}_\mathrm{eff}} = \sigma/L$)
whereas stars of a given group (BSG, YSG or RSG) emits in all wavelengths. The relevant 
comparison to be done here is that the emission at 6000\AA ~have a ${\cal N}_\mathrm{eff}$ lower
than the number of RSG. The relative dispersion at such wavelengths is
around 20\% in the worse case; higher than the dispersion in the UV which does not reach 4\%.

Our mass estimate for NGC~2011 is compatible with the dynamical masses obtained by \cite{KKM93}. On the other hand, 
all masses are a factor 100 higher than the estimate given by \cite{asad12,VGetal2017}, and a factor 10 higher than the 
estimates by \cite{Popescu2012}. There are several possible reason for this discrepancy, which are related with the
use of optical bands to obtain mass inferences as used by \cite{asad12,VGetal2017} and \cite{Popescu2012} analysis. The UV flux decreases monotonically with time, so, for a fixed flux, the inferred mass increases with age.
However, when optical colours are used to constrain the mass, there are always two possible solutions since colours have a peak around 10 Myr (see Sect.  \ref{Sec_indices}), so there is an uncertainty in the age assignation. In addition, the optical flux level has a non-monotonic variation so the mass estimate is strongly linked with the age estimate (which is also linked to the $E(B-V)$ estimate). In the case of stochastic sampling for a given mass, optical bands are more affected by stochastic sampling than the UV continuum, which in practical terms implies that optical bands are less reliable for mass inferences than UV ones. Another possible reason is that the data could have use of different apertures, since sampling effects depends on the amount of stars available to perform the analysis included in the resolution element (we note that \cite{Popescu2012} and \cite{VGetal2017} use a similar value for the observed V-band). Unfortunately, \cite{VGetal2017} does not make use of the UV flux in their mass inferences so we cannot evaluate where would be the origin of the discrepancy.

A similar effect regarding sampling issues is present in CMD analysis where the statistics is given by the number of usable stars (i.e. resolved stars without crowding problems) which will be lower than than the actual number of star in the cluster (c.f. \ref{Sec_CMDs}).

\subsection{Colour-Magnitude Diagrams}
\label{Sec_CMDs}

 \begin{figure*}
   \resizebox{150mm}{!}{\includegraphics[]{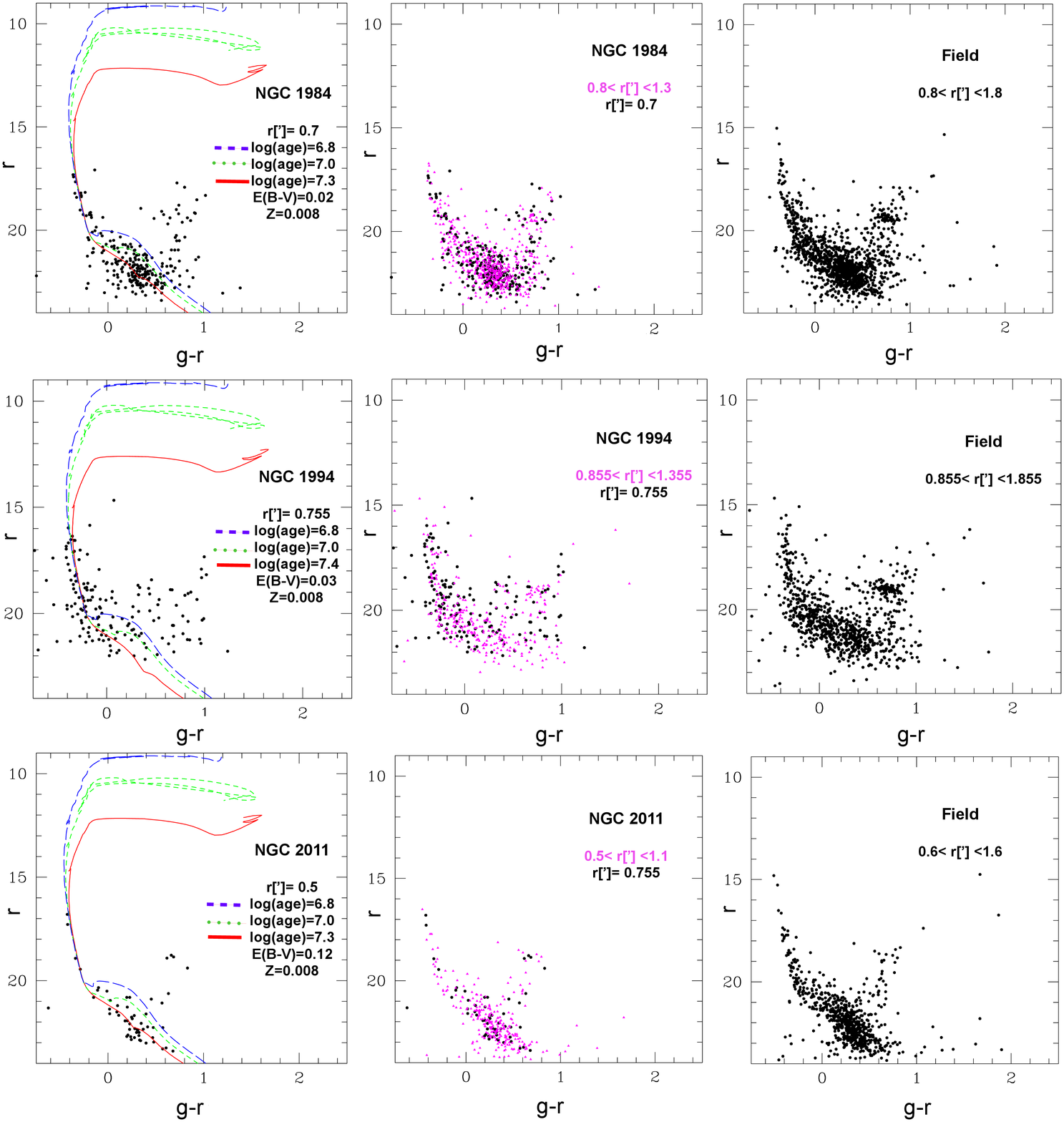}}
\caption[]{g-r vs r CMDs of NGC~1984, NGC~1994 and NGC~2011 with PARSEC isochrones \citep{Bressan12,Chen14,Chen15,Tang14} overlapped. 
 The left panels show the CMDs including only the stars that belong to each cluster within the cluster's radius presented in Table \ref{tab:cmd}. The best fit isochrone in each case (red continuous isochrones) as well as two younger isochrones of $\log t=6.8$ (blue dashed isochrones) and $\log t=7.0$ (green dotted isochrones) are also overlapped. The middle panels show the cluster's stars (black dots) including the stars selected in a 0.50 arcmin concentric annulus located 0.1 outside the cluster's radius (magenta triangles).  
 The right panels show the LMC field stars between 0.1 arcmin and 1.1 arcmin outside the cluster's radius. 
 The optimum-fit isochrones are overlapped with Z=0.008 (see text for more details).}
\label{fig:CMDs}
\end{figure*}

\begin{table}
\caption{CMD results.}
\label{tab:cmd}
\begin{tabular}{lcccc}
\hline
Name &  log(age)  &   E(B-V) & $r$ & $r_{a}$$^{1}$ \\ 
 &    &    &  arcmin & arcmin\\ 
\hline
NGC~1984 & 7.3 & 0.02 & 0.7   & 0.8   $<$r$_{a}$$<$1.8   \\
NGC~1994 & 7.4 & 0.03 & 0.755 & 0.855 $<$r$_{a}$$<$1.855 \\
NGC~2011 & 7.3 & 0.12 & 0.5   & 0.6   $<$r$_{a}$$<$1.6   \\
\hline
\multicolumn{5}{l}{$^1$ Field annulus between $r+0.1$ and $r+1.1$}
\end{tabular}
\end{table}

As an additional cross check for the ages of the clusters, we
produced a suite of {\it g-r} vs {\it r} CMDs.  First, we calculated the mean
apparent diameter of the cluster using the central positions of the cluster from \cite{Bica2008}:

\begin{equation} 
D=(a+b)/2 
\end{equation} 

\noindent where {\it a} is the apparent major axis and {\it b} is the minor
axis. For each cluster we considered those stars within the radius given in
Table~\ref{tab:cmd}  (4th column) as members of the clusters.

Fig.~\ref{fig:CMDs} shows the CMDs for NGC~1984, NGC~1994 and NGC~2011. These are
deep CMDs, reaching down to $\sim$2 magnitudes below the oldest main-sequence
turnoffs.  We fitted PARSEC isochrones \citep{Bressan12,Chen14,Chen15,Tang14} with a metallicity of Z=0.008, appropriate
for young LMC populations \cite[see][]{Carrera08}. See Table~\ref{tab:cmd}
for details.  For each cluster, the optimal-fit isochrone was found by using
different combinations of ages and reddenings using the metallicity constant at Z=0.008.
The CMDs for each cluster within the apparent radius, including the isochrone
fitting, are shown in the left panels of Fig.~\ref{fig:CMDs}. Isochrones of $\log t=6.8$ (blue dashed isochrones), $\log t=7.0$ (green dotted isochrones) are also overlapped in the left panels of Fig.~\ref{fig:CMDs}. 
The age determinations could be severely affected by the LMC field stars' contamination.
In order to avoid this, we followed a similar approach as that presented in
\cite{2010Glatt}.  We sampled the field population within an annulus between
1 and 2 arcmin around each cluster's center outside the apparent radius. These
field populations are shown in the right panels of Fig.~\ref{fig:CMDs}. In order
to estimate the field star contamination, we selected the stars in a 0.50 arcmin
concentric annulus located 0.1 arcmin away from the apparent radius of each
cluster, i.e., between $r+0.1$ arcmin and $r+0.6$ arcmin being $r$ the radius of
each cluster. These stars are plotted as magenta dots in the middle panels of
Fig.~\ref{fig:CMDs}. 

The depth of these CMDs allow us to obtain the ages for all three
clusters. The uncertainties come from differences in the present-day metallicity
of the LMC that, although small, could introduce errors in the age
determination; from small variations in the adopted distance moduli for the LMC;
and from the chosen stellar evolutionary libraries. Taken them together we find
a resulting uncertainty in the age of these clusters of 0.2-0.3 dex.  

To fit the isochrones, we use a constant distance modulus and metallicity and the best-fit isochrones are found by using several combinations of ages and reddening, which we obtained for all three clusters using visual inspection. In the case of the apparent MS turnoffs of the sparse NGC 1994 we use human judgment since there are not enough resolved stars. The log(age)=7.4 isochrones best represent the young Main-Sequence populations in this cluster based in the apparent MS turnoff.

The ages derived using CMD fitting are slightly older than those derived comparing SSPs to the
observed integrated spectra. 
 This discrepancy is not surprising at such young ages due to the differences in the stellar evolutionary tracks, and the fact that although in principle the colour of a stellar population provides a reliable
chronometer to date clusters, in practice, there are, however,  stochastic effects {\cite[see, e.g.][ and discussion in Sect. \ref{Sec_sampling}]{Chiosi88}}. 
\cite{Chiosi06} compared the ages of a large sample of young clusters in the Small Magellanic Cloud obtained with CMD fitting and with integrated light from \cite{Rafelski05} and found a difference of 0.4 dex between the ages of the various clusters. 

\cite{Humphreys1979} observed bright red stars in these clusters. However,  we cannot confirm nor rule out the presence of RSGs due to the fact that stars above 15 mag are saturated in the SMASH imaging.  Having high-resolution images with shorter exposure times could help to further constrain our results in future. 

For this reason, the errorbars for the ages that we obtain from our CMD analysis are not symmetric as we are losing the brightest
stars. Therefore our CMD fitting estimates should be regarded as an upper limit to the age estimates for these clusters.
This underlines the importance of combining multiple observational approaches in this kind of study.

\section{Summary and Discussion}
\label{Sec_discussion}

The $\sim10$ Myr old LMC clusters  NGC~1984, NGC~1994 and NGC~2011 show optical integrated spectra which have strongly concave continua and molecular (TiO) bands far stronger than standard SSP models predict. We  performed an exhaustive exploration of the possible origins of the peculiar appearance of these spectra.  For this purpose, we produced a series of SSP models based on the MILES spectral library which extend to young (6.3 Myr) ages. These models are available on the MILES website\footnote{http://miles.iac.es/}.

We obtained age constraints from the optical spectra using full spectral fitting and the analysis of the absorption and emission Balmer lines, and from the TiO$_1$ spectral index. This was complemented with archival {\it International Ultraviolet Explorer} integrated UV spectra to independently constrain the cluster masses and extinction, and rule out the possibility that stochastic effects are strongly effecting the optical spectra of these clusters; at least in the case of an SSP. We also obtain from the shape of the UV spectra a range of possible ages according different $E(B-V)$ assumptions. In addition, we obtained independent age measurements based upon isochrone fitting of colour-magnitude diagrams. With the younger SSP models and independent mass and age constraints in hand, we explored a number of scenarios in an effort to understand the spectra of these clusters:

{\it Different SSP models:}

To rule out a pathological problem with the MILES model extension to younger ages, we repeated the above analysis with the Geneva-based Granada models. Similarly we used the BC03 models in \citep{Asad16} The results were very similar to those of the MILES models indicating that this is a generic problem with SSP models.

{\it Multiple stellar populations:}

We hypothesized that the clusters contain multiple bursts of star formation, perhaps separated by several Myr. Therefore we made combinations of two-component SSPs and applied our full spectral fitting code. Since the molecular features seen in the optical spectra clearly originate from cool stars, we fixed one component to be at 10~Myr, which maximizes the contribution of RSGs. The second component was allowed to vary freely. We found that combinations of two-component SSPs modestly improved the fit over a single-burst model. 

{\it Top heavy Initial Mass Function:}

We tested whether a top heavy IMF (i.e., an IMF dominated by high mass stars) might affect the optical spectra.
Neither models with a bimodal IMF with $\Gamma_b$ = 0.8, nor an extremely top-heavy IMF ($\Gamma_b$ = 0.3) significantly altered the continuum shapes or affected the molecular bands. This can be understood due to the fact the stellar mass range for stars that contribute significantly to the luminosity is small and therefore shifting the IMF
to higher masses has minimal impact on the SED.

{\it Implementing pre-main sequence contraction:}

We explored whether implementing the long pre-main sequence contraction times of low
mass stars (up to $\sim1$~Gyr for the lowest mass stars) might affect the integrated spectra. This affect was tested by creating ``partial SSPs'' of different mass ranges to mimic the upper and lower stellar ZAMS. We found this makes little difference to the integrated spectra as, at these ages, most of the light contribution comes from very luminous, massive stars. However, this may be important for older ($\sim0.1-1.0$~Gyr) clusters.

{\it Varying the red supergiant contribution:}

In our tests for two-component SSPs, we found that both the continua and molecular features were slightly better reproduced when shifting from LMC to solar metallicity, {\it even though these clusters are known to have LMC metallicities.} The ratio of red- to blue-supergiants is known to increase with increasing metallicity, therefore we reasoned that increasing the contribution of RSG light in the optical spectra, rather than the increase of metallicity {\it per se}, might be at the root of the problem. Therefore we contructed models where we allowed the fraction of RSG to vary over and above that predicted by standard stellar evolution prescriptions. Following this approach, we show that an increase
in flux of $\sim20\%$ from RSGs above the standard prescription can reproduce completely the observed molecular bands, and improve marginally the continuum shapes of the optical spectra.

Based on this work we conclude that the fractional contribution of RSGs in stellar evolution models may require revision in order to improve the prediction of SSP models on the optical spectral range. 
Given the above results we cannot presently rule the possibility of $\sim$Myr age variations in young LMC clusters.

We obtain by our line-strength analysis that cluster ages must be younger than 15 Myr ($\log t = 7.2$) for the three clusters, and older than 8 Myr ($\log t = 6.9$) for NGC~1984 and NGC~2011. Combined with UV analysis, this implies $E(B-V)$ values around 0.1 for these clusters. The ages are consistent with our SED fitting ($\log t \sim = 7$), and isochrone fitting from CMDs (which is $\log t = 7.3 \pm 0.4$ for these two clusters).
However we note that in both fitting techniques, the obtained $E(B-V)$ values is sometimes lower than the minimal value required to explain UV observations, and such underestimation can be related with older ages estimates. 
The case of NGC~1994 is more extreme, since the presence of nebular emission requires an age younger than 8 Myr ($\log t = 6.9$), which implies $E(B-V) > 0.1$ from the UV analysis, whereas best fit SED and CMDs produce ages about or larger than 10 Myr ($\log t = 7$) and $E(B-V)$ values around 0.03 (which lower than $E(B-V)_\mathrm{GAL} = 0.06$), which produce larger ages. Another possibilities not consider in this work are a variable extinction inside the cluster, or a complex star formation history operating at sort time scales which rule out the use of SSP or isochrone fitting since both assume that all stars had been formed at the same time.

By combining different approaches (resolved versus integrated) as well as different wavelength ranges (UV and optical), we obtain 
complementary constraints which should be applied to the different methodologies used in the analysis of stellar clusters. In this work we found that although RSG are not seen in the resolved data, it is clearly seen in the integrated spectra, this is a significant conclusion that emphasizes the importance of the integrated spectra analysis in crowded systems. 
We conclude that fitting in a specific wavelength range might give accurate mathematical results, but it cannot provide the complete physical picture of a given stellar system. Only by  combining multiple approaches  can a complete picture be obtained.

\section*{Acknowledgements}

We thank Lee Patrick, George Meynet, John Beckman and Antonino Milone for the constructive discussions. We thank
Tom\'as Ruiz-Lara for help with the emission correction for NGC~1994.
This work is based on work supported in part by the FRG16-R-10 Grant P.I., R.\ Asa'd from American University of Sharjah. We also acknowledge support from University of Cincinnati while R.\ Asa'd was a summer visiting researcher in summer 2016.
AV and MAB acknowledge support from grant AYA2016-77237-C3-1-P from the Spanish Ministry of Economy and Competitiveness (MINECO). MC acknowledge support from grant AYA2015- 68012-C2-1-P from the Spanish Ministry of Economy and Competitiveness (MINECO). We thank the student AbdulRahman Hachem who helped with some programming parts of the project. 

This research has made use of the NASA/IPAC Extragalactic Database (NED), which is operated by the Jet Propulsion Laboratory, California Institute of Technology, under contract with the National Aeronautics and Space Administration. We also made use of {\sc topcat} software \citep{topcat1,topcat2}.

\bibliographystyle{mnras}

\label{lastpage}
\end{document}